\documentclass [10pt,journal,compsoc]{IEEEtran}
\usepackage{amsbsy,amsmath,amssymb,epsfig,bbm,mathrsfs,multirow,amsthm,setspace,textcomp,xcolor,algpseudocode,mathrsfs,algorithmicx,algorithm}
\ifCLASSOPTIONcompsoc
\usepackage{cite}
\else
\usepackage{cite}
\fi
\usepackage{mleftright}

\usepackage{hyperref}
\usepackage{eurosym}
\newcommand{\LineComment}[1]{\Statex \hfill/*\textit{#1}*/}

\hypersetup{
	colorlinks=true,
	linkcolor=blue,
	filecolor=blue,   
	urlcolor=black,citecolor=blue,}

\newcommand{\lnb}[1]{%
	\ln\mleft(#1\mright)%
}

\newtheorem{theorem}{Theorem}
\newtheorem{lemma}{Lemma}

\DeclareMathAlphabet\mathbfcal{OMS}{cmsy}{b}{n}
\usepackage{array}
\def\BibTeX{{\rm B\kern-.05em{\sc i\kern-.025em b}\kern-.08em
		T\kern-.1667em\lower.7ex\hbox{E}\kern-.125emX}}
\begin{document}
	
	\title{FedChain: Secure Proof-of-Stake-based Framework for Federated-blockchain Systems}
	
	\author{Cong T. Nguyen, Dinh Thai Hoang, Diep N. Nguyen, Yong Xiao, Hoang-Anh Pham, Eryk Dutkiewicz and Nguyen Huynh Tuong %
		\IEEEcompsocitemizethanks{	\IEEEcompsocthanksitem Cong T. Nguyen, Hoang-Anh Pham, and Nguyen Huynh Tuong are with the Ho Chi Minh City University of Technology, VNU-HCM, Vietnam. E-mail:  \{ntcong.sdh19, anhpham\}@hcmut.edu.vn.
			
			\IEEEcompsocthanksitem Diep N. Nguyen, Dinh Thai Hoang, and Eryk Dutkiewicz are with the School of Electrical and Data Engineering, University of Technology Sydney, Australia.	E-mail:  \{diep.nguyen, hoang.dinh, eryk.dutkiewicz\}@uts.edu.au.
			
			\IEEEcompsocthanksitem Yong Xiao is with the School of Electronic Information and Communications, Huazhong University of Science and Technology, Wuhan, China. E-mail: yongxiao@hust.edu.cn
		}
		
		\thanks{}
		\vspace{-5mm}}

	\IEEEtitleabstractindextext{	
		\begin{abstract}
In this paper, we propose FedChain, a novel framework for federated-blockchain systems, to enable effective transferring of tokens between different blockchain networks. Particularly, we first introduce a federated-blockchain system together with a cross-chain transfer protocol to facilitate the secure and decentralized transfer of tokens between chains. We then develop a novel PoS-based consensus mechanism for FedChain, which can satisfy strict security requirements, prevent various blockchain-specific attacks, and achieve a more desirable performance compared to those of other existing consensus mechanisms. Moreover, a Stackelberg game model is developed to examine and address the problem of centralization in the FedChain system. Furthermore, the game model can enhance the security and performance of FedChain. By analyzing interactions between the stakeholders and chain operators, we can prove the uniqueness of the Stackelberg equilibrium and find the exact formula for this equilibrium. These results are especially important for the stakeholders to determine their best investment strategies and for the chain operators to design the optimal policy to maximize their benefits and security protection for FedChain. Simulations results then clearly show that the FedChain framework can help stakeholders to maximize their profits and the chain operators to design appropriate parameters to enhance
FedChain’s security and performance.

		\end{abstract}
		
		\begin{IEEEkeywords}
			Blockchain, Proof-of-Stake, cross-chain transfer, sidechain, multiple-blockchain, and Stackelberg game.
	\end{IEEEkeywords}}
	
	\maketitle
	\IEEEdisplaynontitleabstractindextext
	
	\IEEEpeerreviewmaketitle

	\thispagestyle{empty}
	\section{Introduction}
	\subsection{Motivation}
	Over the last few years, the development of the blockchain technology has attracted massive attention. A blockchain is an append-only ledger of transactions shared among the participants in a peer-to-peer network. With the help of consensus mechanisms, once a transaction enters the blockchain, it cannot be changed without the consensus of the majority of the network. Beside data immutability, the consensus mechanism also plays a key role in ensuring that such a decentralized network can reach the consensus without a central authority, thereby avoiding the single-point-of-failure. Moreover, advanced cryptography techniques such as digital signatures and asymmetric keys~\cite{wangsurvey,Xiaosurvey} enable blockchain users to create easily verifiable but impossible to forge proofs of authentication for assets (i.e., blockchain tokens) while enhancing the anonymity and privacy of users. As a result, blockchain can enable trusted transactions among network participants even in an open and decentralized environment. With such outstanding benefits, blockchain has been implemented as the backbone of numerous applications in many areas such as finance, healthcare, and Internet-of-Things (IoT)~\cite{wangsurvey,Xiaosurvey}. 
	
	Despite its popularity and potential, blockchain has been facing various challenges. The rapid development of blockchain and the massive popularity of cryptocurrency have lead to the creation of a plethora of blockchain networks. For example, the number of cryptocurrency networks has increased nearly four times in just one year (from 2000 cryptocurrencies in 2019 to 7400 by the time this article is written, i.e., December 2020~\cite{coinmarketcap}). These blockchain networks are currently employing diverse consensus mechanisms, which results in severe fragmentation since these networks cannot communicate with each other. However, there are many blockchain-based applications where the ability to transfer assets between different blockchains is essential, such as coalition loyalty programs and retail payment. For example, in coalition loyalty programs, users need to exchange their loyalty points among different programs, which are stored on different blockchains in the forms of blockchain tokens. Similarly, in retail payment, vendors might only accept a certain type of tokens, and thus the users need to exchange their tokens to another type. However, for single blockchain networks, users who want to exchange tokens have to rely on trusted centralized exchange platforms, e.g., Binance~\cite{Binance} and Kraken~\cite{Kraken}, which is against the decentralized nature of blockchain and poses serious security threats. Particularly, there have been many attacks on these exchanges, resulting in a cumulative loss of more than \$1 billion~\cite{hack} over the last few years. Moreover, the trade-off between performance and security in consensus mechanism designs usually leads to high delay and low processing throughput. For example, Bitcoin needs 1 hour to confirm a transaction and can only process less than 7 transactions per second~\cite{Xiaosurvey}, which hinders blockchain applicability in many scenarios. Thus, this necessitates an effective framework that not only allows the interoperability among blockchains networks, but also guarantees the security and performance of each individual network.
	
	To address these problems, the sidechain technology~\cite{side} has been developed to enable the formation of the federated-blockchain system, which consists of multiple blockchains to allow the users to transfer assets to any blockchain within. The core mechanism of the sidechain technology that enables the exchanging of tokens between different chains is the two-way peg mechanism. Specifically, when a user wants to transfer its assets from one chain to another chain, the user first creates a transaction to lock its assets on the originating chain. Then, a group of validators, selected by the two-way peg mechanism, will verify and confirm this transaction, and create a corresponding amount of assets on the destination chain. Typically, the two-way peg mechanism can ensure that the cross-chain transfers are secure and cannot be reverted, i.e., avoid cross-chain double-spending attacks~\cite{side}. However, the development of the sidechain technology is still in a nascent stage, and it does not fully satisfy the security nor the performance requirements of federated-blockchain systems. Particularly, the ability to transfer assets between multiple chains may lead to centralization to a single chain, e.g., mining power centralization in Proof-of-Work (PoW) and stakes centralization in Proof-of-Stake (PoS). This poses a security threat to the other chains in the same federation. Moreover, most current sidechain applications still employ the PoW mechanism which requires huge energy consumption and has very low processing capabilities~\cite{PoS,wangsurvey}. Therefore, a secure and effective framework, which can address both security and performance issues for cross-chain transfers, is in urgent need for the future development of blockchain networks. 
	

	\subsection{Related Work}
	Sidechain technology was first introduced in~\cite{side} as a novel method to facilitate cross-chain transfers. Particularly, two-way peg and Simplified Payment Verification (SPV) proof mechanisms are developed so that the validators can verify and confirm transactions between different blockchains. Although this work paves the way for many research works and applications, the security and performance issues of sidechain are only briefly mentioned and not well investigated~\cite{side}. After the introduction of the sidechain technology, there have been several notable real-world applications such as PoA~\cite{PoA}, Liquid~\cite{Liquid}, and RSK~\cite{RSK}. However, these applications are facing several challenges. In particular, the PoA approach relies on a fixed federation of 23 validators to validate the cross-chain transactions between the Ethereum~\cite{Ethereum} and several sidechains. This results in a low decentralization level for the consensus process. Moreover, these validators' identities are publicly known, making them easier to be targeted by attackers. Similarly, the Liquid approach~\cite{Liquid} also relies on a federation to validate cross-chain transactions. Although these validators are not publicly known, they are chosen only by the network operators, and thus Liquid is not a public blockchain network. Moreover, Liquid is using a version of the PoW consensus mechanism which requires even more computational resources than Bitcoin (Liquid requires the validators to run a Bitcoin node in parallel with a Liquid node). Similar to Liquid, RSK employs a federation to validate transactions via a PoW-based mechanism. Although RSK is more decentralized, i.e., the federation in RSK is determined by public voting, RSK is still limited by the huge energy consumption of the PoW mechanism.
	
	Different from the PoW mechanism, the PoS mechanism enables the blockchain participants to reach the consensus by proving tokens ownership. As a result, the PoS mechanism is much more energy-efficient and can achieve higher transaction processing speed compared to those of the PoW mechanism~\cite{wangsurvey,PoS,Xiaosurvey}. Due to those advantages, recent research works in the area of the sidechain technology have shifted towards the PoS mechanism. In~\cite{side1}, a cross-chain transfer protocol is developed for cross-chain transfers between a primary blockchain (main chain) and a secondary chain (sidechain). To validate the cross-chain transactions, the protocol relies on a set of certifiers who are chosen by the main chain. A major advantage of the proposed protocol is the independence between the side chain and main chain in terms of security and operations. However, the security of this protocol is not analyzed. In~\cite{side2}, the authors propose a sidechain system, in which both the sidechain and the main chain employ a PoS mechanism, i.e., Ouroboros. Unlike the previous works, this work focuses more on the security aspects of the sidechain technology, providing formal definitions and robust security analyses. Moreover, the proposed system ensures the independence in terms of security between the blockchains within. However, the risk of centralization is not addressed. Similar to~\cite{side1}, the authors in~\cite{side3} also introduces a cross-chain transfer protocol to allow interoperability between a main chain and a side chain. The cross-chain transfer protocol in~\cite{side3} is proposed with formal definitions, and a consensus mechanism is also presented in a similar way as in~\cite{side2}. However, there are several limitations in this work, such as the lack of formal security analysis and the unaddressed risk of centralization. 
	
	To the best of our knowledge, the risk of centralization in federated-blockchain systems has not been addressed in any previous work. Specifically, the ability to transfer tokens between blockchains may lead to situations where the users centralize to a single blockchain in the system. Particularly, in PoS blockchains, a user who participates in the consensus mechanism has a chance to be selected to create new blocks and obtain a reward. That chance is directly proportional to the tokens the user possesses in the network~\cite{wangsurvey,PoS}. Therefore, a blockchain with a higher reward might attract more users and tokens, as the users will transfer their tokens to that blockchain to earn more profits. Such centralization of tokens and users may have negative impacts on the security and performance of the other blockchains in the same system. This is because the state of each PoS blockchain is determined by the majority of stakes (tokens), i.e., users who have more stakes (tokens) are more likely to be selected to add new blocks. Consequently, it is easier for attackers to target the blockchains that have fewer tokens. This can significantly impact these blockchains' security and performance. Furthermore, since the cross-chain transfer requires the confirmation of transactions in both the originating and destination chains, the centralization of stakes also reduces the overall system performance. More detailed analysis of these negative impacts will be presented in Section~\ref{sec:con}.

	\subsection{Contributions and Paper Organization}
	The main contributions of this paper are summarized as follows:
	\begin{itemize}
		\item Propose FedChain, an effective and secure framework for cross-chain transfer in federated-blockchain systems. Particularly, Fedchain facilitates two-way transfers of assets between different blockchains in the system by utilizing the sidechain technology. Moreover, to address the security and performance limitations of current sidechain technology, we develop a PoS consensus mechanism and a Stackelberg game model specifically for FedChain.
		\item Develop a novel PoS-based consensus mechanism for the individual blockchain in FedChain. By designing new effective rules, we can significantly improve the security and performance of the consensus mechanism. Particularly, through theoretical and numerical analyses, we prove that the proposed consensus mechanism can satisfy the persistence and liveness properties~\cite{sec}, prevent many blockchain-specific attacks, and achieve a more desirable transaction confirmation time compared to several other mechanisms such as the Nakamoto protocol (of Bitcoin)~\cite{Bitcoin} and Ouroboros (of Cardano)~\cite{Ouroboros}. 
		\item  We develop an incentive mechanism using a Stackelberg game model~\cite{game} for FedChain in order to provide additional benefits for the users, enhance FedChain's security and performance, and address the problem of centralization in the sidechain technology. Moreover, we propose a highly effective utility function for the chain operators, which can help to attract more stakes to the individual chains while still ensuring the overall decentralization of the system. To the best of our knowledge, this is the first paper addressing the risk of centralization in federated-blockchain systems. Furthermore, by analyzing interactions between the stakeholders and the chain operators, we can prove the uniqueness of the Stackelberg equilibrium and find the exact formula for this equilibrium. These results are especially important for the stakeholders to determine their best investment strategies and for the chain operators to design the optimal policy to enhance the system's security and transaction processing capabilities.
		\item  Extensive simulations are performed to evaluate the system performance of FedChain. The simulation results then confirm the analytical results and show that FedChain can help the users to maximize their profit and the blockchain operators to determine their optimal blockchain parameters to improve the system's security and performance. 
	\end{itemize}
	
	The rest of this paper is organized as follows. We first present the federated-blockchain framework in Section~\ref{sec:SM}. We then analyze the proposed consensus mechanism for our framework in Section~\ref{sec:con}. After that, we introduce and analyze the Stackelberg game in Section~\ref{Stackel}. Finally, simulations and numerical results are presented in Section~\ref{sim}, and conclusions are drawn in
	Section~\ref{sec:Sum}.

	\section{Federated-blockchain System}
	\label{sec:SM}
	\subsection{System Overview}
\begin{figure*}[!t]
 	\includegraphics[width=1.3\columnwidth]{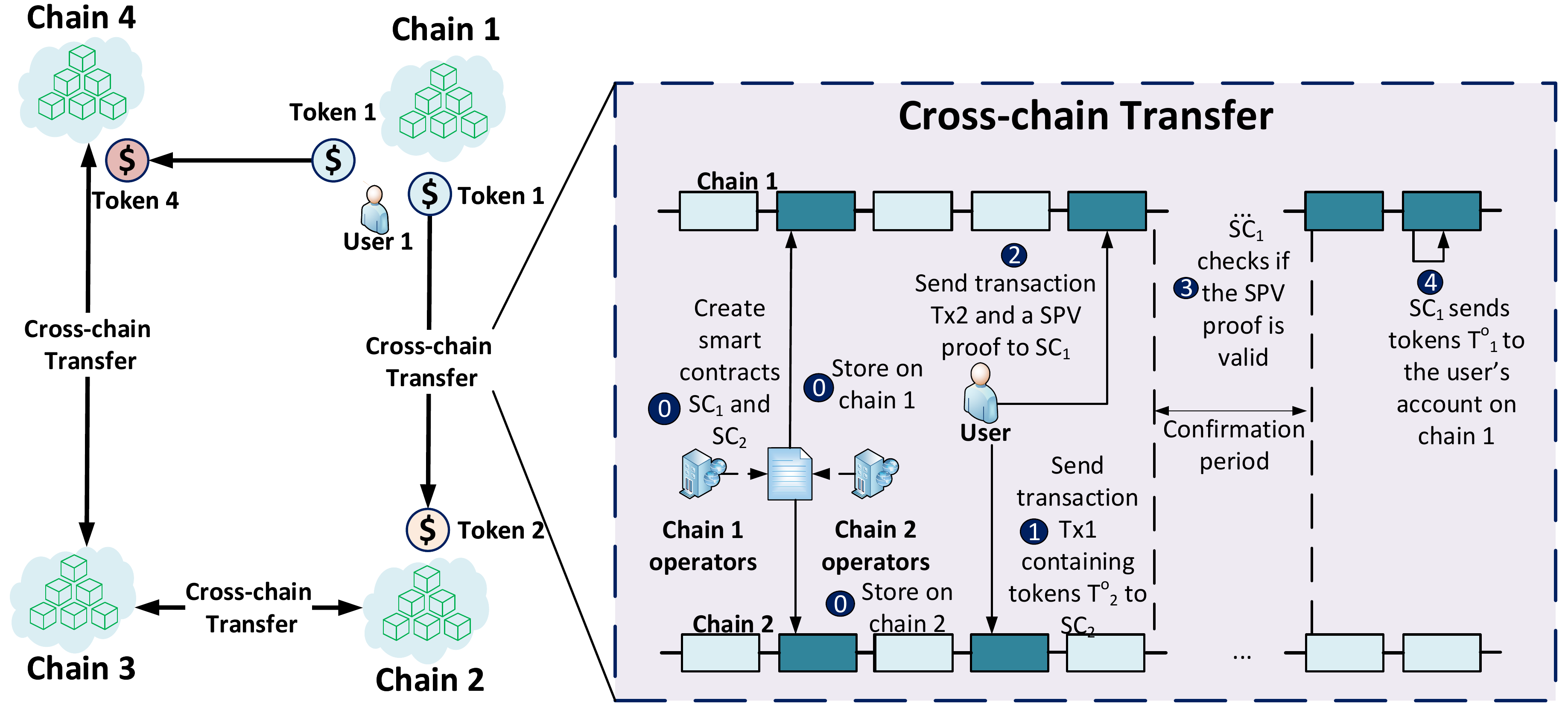}
 	\centering
 	\caption{The federated-blockchain system.}
 	\label{Fig:system}
 \end{figure*}

	Before elaborating on our proposed consensus mechanism and incentive mechanism, we provide a brief overview of the federated-blockchain system and the cross-chain transfer procedure in this section~\cite{side,sidesurvey}. As illustrated in Fig.~\ref{Fig:system}, the system is composed of two types of entities as follows:
	\begin{itemize}
		\item \textbf{Chains (blockchains):} In FedChain, individual blockchain networks, managed by blockchain operators, can communicate with each other via the cross-chain transfer protocol. Each chain has its own type of token and an individual consensus mechanism. When a new blockchain network wants to join the system, it only needs to negotiate with the existing chains and create smart contracts accordingly.
		\item \textbf{Users:} Users are the participants of the chains in the system. These users can freely exchange different types of tokens by using the smart contracts created by the operators. They can also participate in the consensus mechanism in every chain to earn economic profits through block rewards. 
	\end{itemize} 

\subsection{Cross-chain Transfer Procedure}		
	
	The SPV mechanism allows tokens from one chain to be securely transferred to another at a predetermined rate. When a user wants to prove that a transfer transaction from an originating chain to a destination chain is valid, an SPV proof is submitted. This proof shows that the transfer transaction belongs to a valid block of the originating chain. Although this process takes a long time for confirmation, it eliminates the risk of centralization and single-point-of-failure compared to those of the centralized and federated scheme~\cite{sidesurvey}. Therefore, the SPV proof is selected as the cross-chain transfer mechanism in our proposed FedChain. As illustrated in Fig.~\ref{Fig:system}, the SPV-based token exchange procedure consists of several steps as follows:
	\begin{itemize}
		\item \textit{Step 0:} Two chains negotiate an agreement which specifies the exchange rate between the two tokens. The chain operators then create in each chain a smart contract according to the agreement. 
		\item \textit{Step 1:} When a user wants to exchange $T^o_2$ tokens into $T^o_1$ tokens, the user sends a transaction $\rm Tx1$, containing $T^o_2$ tokens, from its account on chain 2 to the smart contract $\rm SC_2$.
		\item \textit{Step 2:} The user then sends a transaction $\rm Tx2$ and an SPV proof from its account on chain 1 to $\rm SC_1$. $\rm Tx2$ then triggers $\rm SC_1$ to validate the SPV proof.
		\item \textit{Step 3:} During the confirmation period, $\rm SC_1$ checks (1) the validation of the SPV proof and (2) any conflicts of the submitted SPV proof.     
		\item \textit{Step 4:} After the confirmation period, $\rm SC_1$ sends a number of $T^o_1$ tokens to the customer's address on chain 1 in accordance with the exchange rate.
	\end{itemize} 

	The security features of the SPV proof mechanism are proven in~\cite{side}. Generally, the SPV proof points to the block that contains the cross-chain transfer transaction in the originating chain. Therefore, the validators only have to validate the block that contains the transaction. Thus, the security of the SPV proof only relies on the security of the originating chain, i.e., the SPV proof is secure if the originating chain is secure. However, this leads to a drawback of the SPV proof mechanism, which is the low confirmation speed (the validators have to wait until the transaction is confirmed on the originating chain). Moreover, as the stakes can be transferred between chains, if the security of one chain is violated, the whole system will fail. Therefore, in the next section, we will propose an effective consensus mechanism that can achieve lower transaction confirmation time compared to other conventional mechanisms while satisfying the persistence and liveness properties~\cite{sec} and being able to prevent various blockchain attacks.
\section{FedChain's Consensus Mechanism}
	\label{sec:con}
	In this section, we develop an effective consensus mechanism for FedChain with four new consensus rules based on the consensus mechanism proposed in~\cite{Ouroboros}. Compared with other conventional consensus mechanisms such as~\cite{PoA2,CoA,Tendermint,Algorand,Ouroboros,Casper,snow}, our proposed consensus mechanism can satisfy both the liveness and persistence properties, prevent various blockchain attacks, and achieve an especially low transaction confirmation time as discussed in the following.

\subsection{Proposed Consensus Mechanism}
	\subsubsection{Epochs and time slots}
\begin{figure}[!]
		\includegraphics[width=\columnwidth]{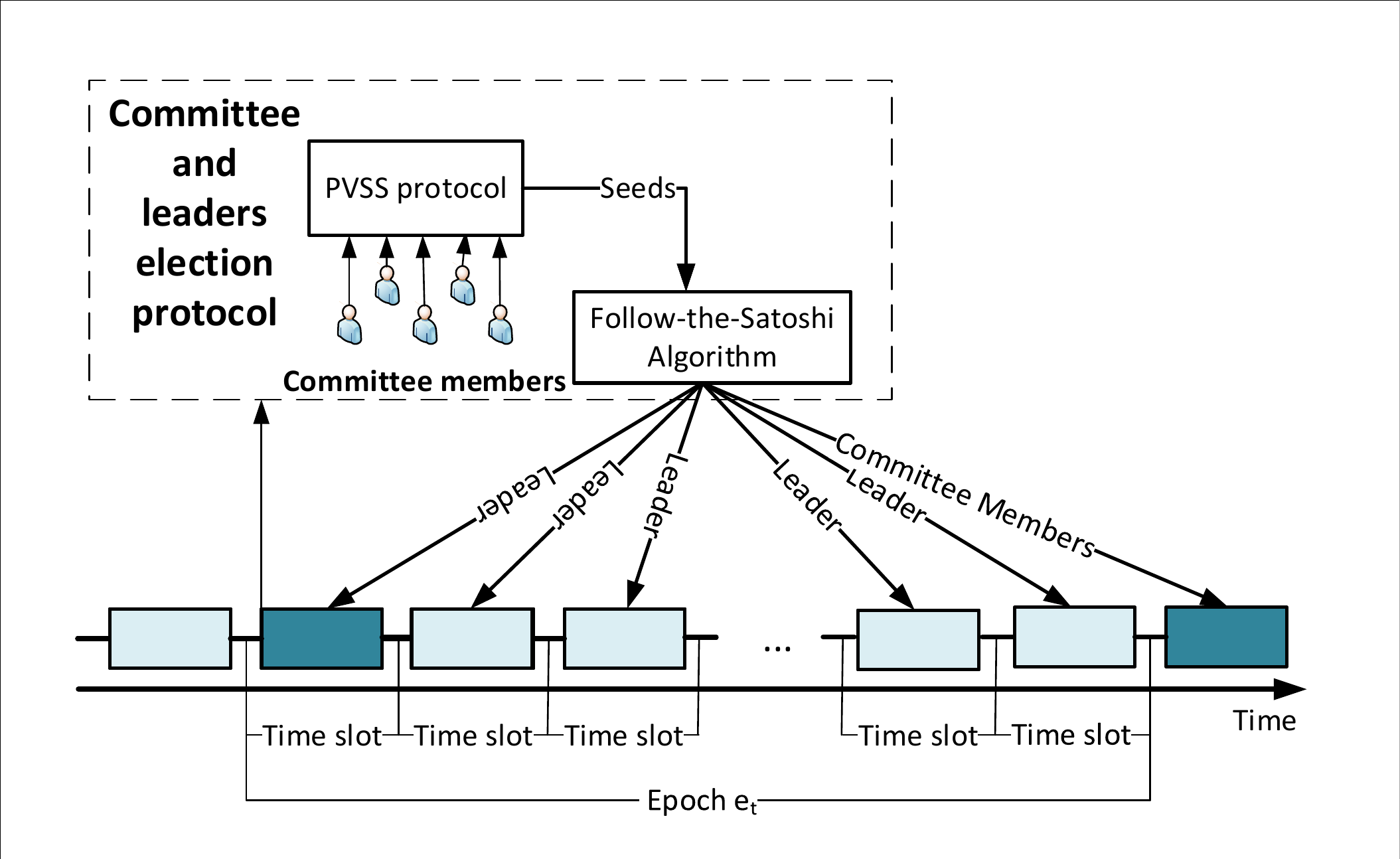}
		\centering
		\caption{Epoch-based committee and leader election.}
		\label{Fig:epoch}
	\end{figure}

	As illustrated in Fig.~\ref{Fig:epoch}, time is divided into epochs, and each epoch is divided into time slots in FedChain's consensus mechanism. At the first time slot of epoch $e_k$, a committee consisting of some users (stakeholders) executes an election protocol to elect the leaders for the epoch $e_k$, such that for each time slot there is one designated leader who adds one new block to the chain. Similar to~\cite{Ouroboros}, we assume that a time slot duration of 20 seconds is sufficient for the leader to broadcast a block to every node in the chain. The committee also select the committee members for the epoch $e_{k+1}$.
	\subsubsection{Leaders and committee election protocol}
	To elect the leaders and committee, the current epoch's committee members execute the Publicly Verifiable Secret Sharing (PVSS) protocol~\cite{PVSS} to create seeds for the Follow-the-Satoshi (FTS) algorithm~\cite{PoS}. The PVSS protocol allows the participants to produce unbiased randomness in the form of strings and any network user to verify these strings, as long as the majority (51\%) of participants are honest, as proven in~\cite{PVSS}. Once the random strings are created, they are used as the seeds for the FTS algorithm. The FTS algorithm is a hash function that takes any string as input and outputs token indices~\cite{PoS}. The current owners of these tokens are then chosen as the leaders of this epoch or committee members of the next epoch.  
	
	The probability $P_n$ that user $n$ is selected to be the leader and committee member by the FTS algorithm in a network of $N$ stakeholders is
	\begin{equation}
		\label{eq:PoS}
		P_n=\dfrac{s_n}{\sum_{i=1}^{N}s_i},
	\end{equation}
	where $s_n$ is the number of stakes (tokens) of stakeholder $n$. As observed in~\eqref{eq:PoS}, the more stakes a stakeholder has, the higher chance it can be selected to be the leader. Compared to~\cite{Ouroboros}, we design four new consensus rules as follow:
	\begin{itemize}
\item $I_1$: After executing the PVSS protocol, the leader list is broadcast to every node in the chain.
\item $I_2$: If a leader fails to broadcast its block during its designated time slot (e.g., being offline during its time slot), an empty block will be added to the chain
\item $I_3$: Once a block is broadcast, the designated leader will not change the block at any later time.
\item $I_4$: Upon receiving two forks (different versions of the chains), an honest user will adopt the longest valid fork, i.e., the longest fork that has no conflicting blocks and each block is signed by a designated leader.
	\end{itemize} 
These new consensus rules help to considerably reduce the probability that an adversary can successfully create an alternative version of the chain, thereby significantly improving the chain's security and performance. The detailed analysis will be discussed in Theorem 1.
	\subsubsection{Incentive mechanism}
	The incentive mechanism plays a crucial role in ensuring that the stakeholders follow the consensus mechanism properly. To this end, the incentive mechanism needs to incentivize consensus participants via a reward scheme and penalize malicious behavior via a penalty scheme.
	
	For the reward scheme, a leader will receive a fixed number of tokens when the leader adds a new block to the chain. This is also to incentivize the leaders to be online during their designated time slots. In single-blockchain settings such as Bitcoin~\cite{Bitcoin} and Cardano~\cite{Ouroboros}, the block reward is set at a fixed value for a long period of time, e.g., 4 years in Bitcoin. However, in FedChain, having a fixed block reward scheme may pose security threats. The reason is that the stakes can be transferred between chains in our system, and the total network stakes can also vary in times, e.g., stakes increase from block rewards, and the stakes decrease from cross-chain transfers, etc. Since the probability that a stakeholder is elected to be the leader and able to obtain a block reward depends on the individual chain's stakes, stakeholders may transfer their stakes to a chain with a higher block reward to earn more profits. Consequently, this may attract stakes into a single chain and make it easier for adversaries to control the majority of stakes in the other chains. Therefore, in the following sections, we analyze the stakeholder rational strategy and propose a dynamic reward scheme to protect the decentralization of the whole system. With our proposed dynamic reward scheme, at the end of each epoch, the chains will adjust new block reward values for the next epoch, taking the total network stakes and the final stakes distribution among the chains in the current epoch into account. The dynamic reward scheme will be discussed in more details in Section~\ref{Stackel}.
	
	For the penalty scheme, the leader is required to make a deposit that will be locked during its designated epoch to prevent nothing-at-stake, bribe~\cite{PoS}, and transaction denial attacks~\cite{Ouroboros}. The stakes of committee members are also locked during the epoch that they are serving in the committee to prevent long-range attacks~\cite{PoS}. How the proposed penalty scheme can prevent the mentioned attacks will be discussed in the following security analysis.

\subsection{Security Analysis}
\subsubsection{Adversary and attack models}
Since the SPV proof mechanism's security depends on the security of the individual chains, the security of the whole system also relies on the security of each chain.
We consider two types of adversaries that target the individual chains, aiming to perform attacks such as double-spending, grinding, nothing-at-stakes, bribe, transaction denial, and long-range attacks~\cite{PoS}. As illustrated in Fig.~\ref{Fig:adver}, the considered types of adversaries are:
	\begin{figure*}[!]
	\includegraphics[width=\textwidth]{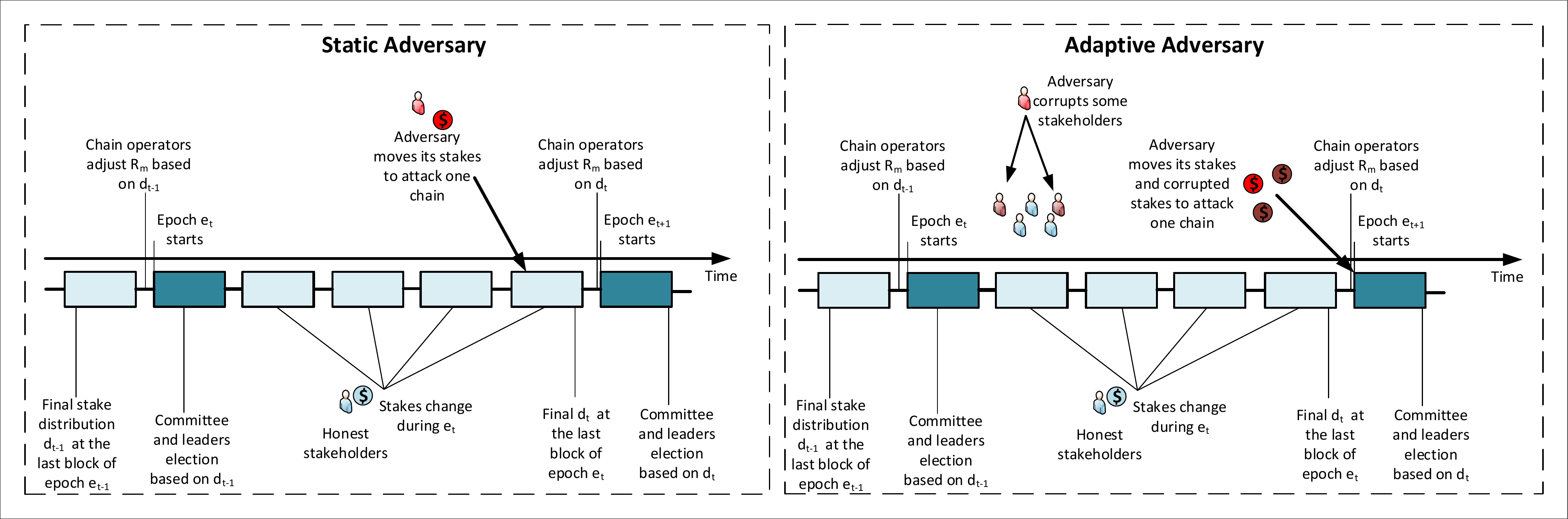}
	\centering
	\caption{Illustrations of the considered adversaries.}
	\label{Fig:adver}
\end{figure*}
\begin{itemize}
		\item{\textbf{Static Adversary:}} This type of adversary uses a stake budget $B_A$ to attack a chain. Let $B_n$ and $\gamma$ denote the stake budgets of stakeholder $n$ and the honest stake ratio, respectively. Then, the ratio of adversarial stakes is $1-\gamma=\dfrac{B_A}{\sum_{n=1}^N B_n + B_A}$. 
		
		\item{\textbf{Adaptive Adversary:}} In contrast to the static adversary setting, the adaptive adversary does not have a fixed number of stakes. However, this type of adversary can choose to corrupt $N_A$ honest stakeholders and use their stakes to attack. Let $\mathcal{N}_A$ denote the set of corrupted stakeholders, the budget of the adaptive adversary can be defined by $B_{A}=\sum_{i \in \mathcal{N}_{A}} B_i$.
\end{itemize}

The models for the blockchain-specific attacks considered in this paper are as follows:
\begin{itemize}
	\item{\textbf{Double-spending attack:}} For such kind of attack, the attacker aims to revert a transaction that has been confirmed by the network (to gain back the tokens it has already spent). First, the attacker creates a transaction $\rm Tx1$ in block $\mathcal{B}_i$ and waits until the block is confirmed. Then, the attacker can either create a conflicting transaction $\rm Tx2$ or erase the block $\mathcal{B}_i$ from the chain, so that the proof of its spending is gone.
	
	\item{\textbf{Grinding attack:}} In grinding attacks, the attacker attempts to influence the leader and committee election protocol to unfairly increase its chance to be selected as a leader or a committee member. Generally, in protocols where the seeds of the FTS algorithm are derived from the block header, the attacker can check many possible different block contents (because block headers are created by hashing the block contents) to determine which one can give the attacker the best chance to be elected as a leader again.
	
	\item{\textbf{Nothing-at-stake attacks:}} This type of attack specifically targets the PoS blockchains because, in contrast to PoW, blocks in PoS can be created with very little computation. In this attack, the attacker tries to create many forks or conflicting transactions. For example, the attacker can create two transactions to spend the same tokens at two vendors, i.e., $\rm Tx1$ in fork $\mathcal{C}_1$ and $\rm Tx1$ in fork $\mathcal{C}_2$. At this point, although both the transactions are not \textit{confirmed}, they are both \textit{valid} (not conflicted within their own fork). 
	
	\item{\textbf{Bribe attacks:}} For such attacks, the attacker tries to bribe the leaders to create specific blocks, e.g., to support other types of attacks such as double-spending or transaction denial.
	
	\item{\textbf{Transaction denial attack:}} In this attack, the attacker tries to prevent transactions of every or some specific users from being included in the chain. To achieve this objective, the attacker has to either block the users' connection to the blockchain or not include the transactions when the attacker is the leader.  
	
	\item{\textbf{Long-range attack:}} In a long-range attack, a leader immediately transfers its stakes to another account at the beginning of its designated epoch, and thus it can behave maliciously, e.g., performing attacks, for the rest of the epoch without consequences.
\end{itemize}

\subsubsection{Blockchain properties}
To maintain the blockchain's security, a consensus mechanism must satisfy the following properties~\cite{sec}: 
\begin{itemize}
	\item \textbf{Persistence:} Once a transaction is confirmed by an honest user, all other honest users will also confirm that transaction, and the transaction's position in the blockchain is the same for all honest users. 
	\item \textbf{Liveness:} After a sufficient period, a valid transaction will be confirmed by all the honest users.
\end{itemize}
In FedChain, persistence ensures that once a transaction is confirmed, it cannot be reverted. Without the persistence property, the adversary can successfully perform a double-spending attack by firstly sending a transaction to spend some tokens. After that transaction is confirmed, the adversary can create a fork to erase the transaction from the blockchain. If that fork is accepted by the honest users, the adversary can gain back the tokens it already spent. While the persistence property ensures data immutability, the liveness property ensures that every valid transaction will eventually be included in the chain. Without liveness, an attacker can block every transaction in a blockchain. The persistence and liveness properties are ensured if the consensus mechanism satisfies the following properties~\cite{sec}:
\begin{itemize}
	\item \textbf{Common prefix (CP) with parameter $\kappa \in \mathbb{N}$:} For any pair of honest users, their versions of the chain $\mathcal{C}_1,\mathcal{C}_2$ must share a common prefix. Specifically, assuming that $\mathcal{C}_2$ is longer than $\mathcal{C}_1$, removing $\kappa$ last blocks of $\mathcal{C}_1$ results in the prefix of $\mathcal{C}_2$. 
	\item \textbf{Chain growth (CG) with parameter $\varsigma \in \mathbb{N}$ and $\tau\in (0,1]$:} A chain possessed by an honest user at time $t+\varsigma$ will be at least $\varsigma\tau$ blocks longer than the chain it possesses at time $t$.
	\item \textbf{Chain quality (CQ) with parameter $l \in \mathbb{N}$ and $\mu \in (0,1]$:} Consider any part of the chain that has at least $l$ blocks, the ratio of blocks created by the adversary is at most $1-\mu$. In the ideal case, $1-\mu$ equals the adversarial ratio $1-\gamma$.
\end{itemize}

Let $\rm Pr_{CP}$, $\rm Pr_{CG}$, and $\rm Pr_{CQ}$ denote the probabilities that the CP, CG, and CQ properties are violated. We prove that FedChain's consensus mechanism can satisfy the CP, CG, and CQ properties with overwhelming probability, i.e., $\rm Pr_{CP}$, $\rm Pr_{CG}$, and $\rm Pr_{CQ}$ are overwhelmingly low ($<0.1\%$), in the following Theorem.

\begin{theorem}
	FedChain's consensus mechanism can satisfy the CP, CG, and CQ properties with overwhelming probabilities.
\end{theorem}
\begin{IEEEproof}
	See Appendix~\ref{proof1}
\end{IEEEproof}
Fig.~\ref{Fig:prob} illustrates the CP and CQ violation probabilities under different parameter values. As the adversarial ratio increases (i.e., the adversary controls more stakes in the chain), the attacker has more chances to successfully attack. However, the higher $\kappa$ is, the lower the CP violation probability is. This means that the longer since a transaction is added to the chain, the more stable the transaction becomes. For example, if a transaction is at least seven blocks deep in the chain, the adversary has less than 1\% chance to revert it, even if the adversary controls nearly 50\% of the total network stakes. In contrast, if the transaction is only four blocks deep, the adversary with 49\% stakes has more than 5\% chance to revert the transaction. This implies that the more stakes the adversary controls, the longer it takes to confirm a transaction, which is directly related to the performance and security of the chain. 

For the $\rm{Pr}_{CQ}$, the more blocks we consider, the higher chance the adversary can create more than $(1-\gamma)l$ blocks. For example, an adversary controlling 30\% of network stakes has less than 0.1\% chance to create more than three in ten blocks, but it has around 0.3\% chance to create more than 30 in 100 blocks. This could be harmful to the network if the adversary wants to reduce the network's throughput (i.e., blocks/time slot). For example, an adversary with 30\% network stakes has 0.3\% chance to reduce the network throughput by 30\% during 100 time slots by creating only empty blocks every time it is elected to be the leader. 
	\begin{figure}[!]
	\includegraphics[width=\columnwidth]{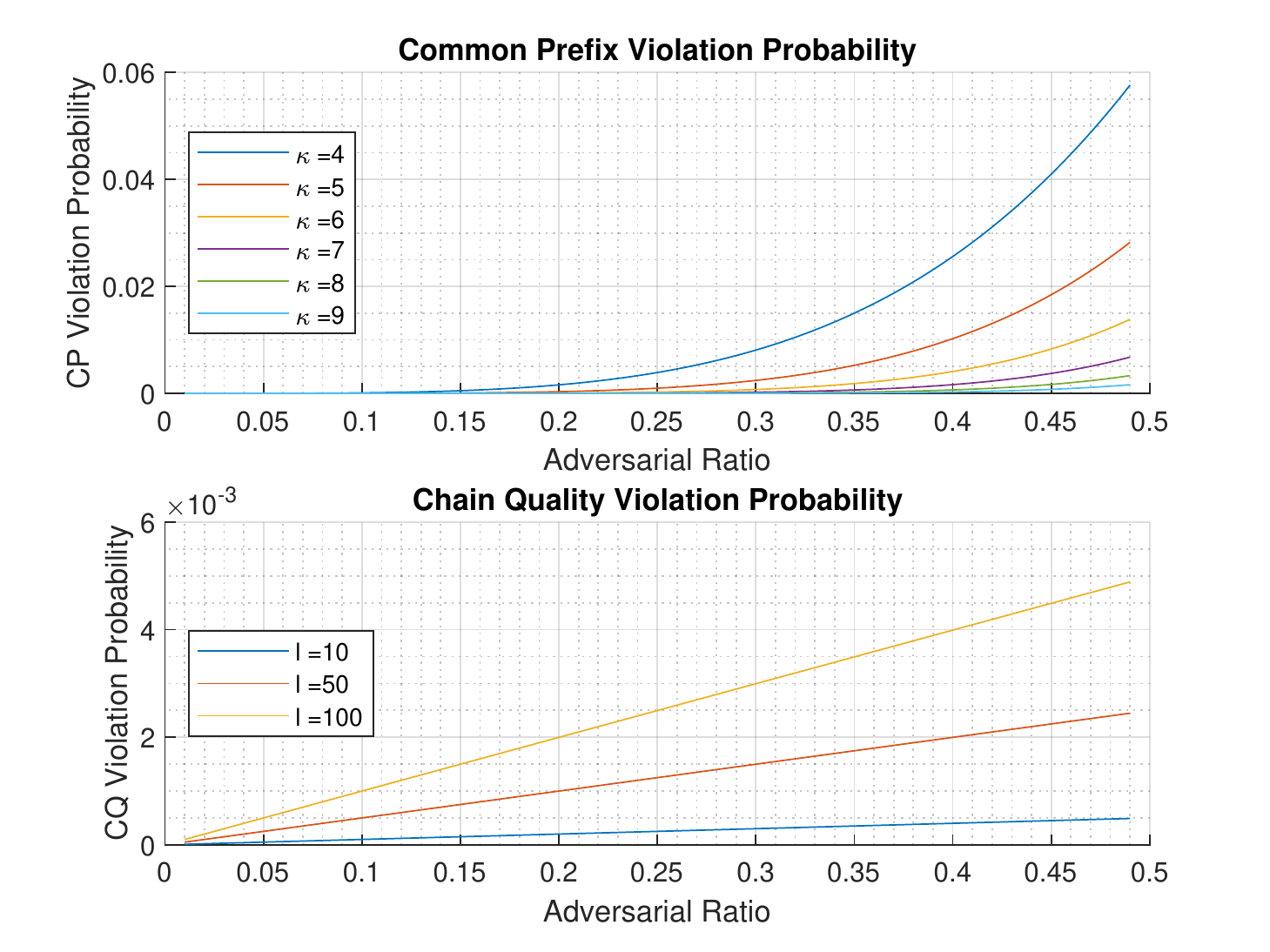}
	\centering
	\caption{Blockchain properties violation probabilities.}
	\label{Fig:prob}
\end{figure}
\subsubsection{Blockchain attacks prevention}
\label{attack}
In the following Theorem, we prove that our FedChain's consensus mechanism is able to prevent a variety of emerging blockchain attacks such as double spending, grinding, bribe, nothing-at-stakes, and long-range attacks. 
\begin{theorem}
	FedChain's consensus mechanism can prevent double-spending, nothing-at-stakes, bribe, transaction denial attacks, grinding, and long-range attacks according to the considered adversary models.
\end{theorem}
\begin{IEEEproof}
		See Appendix~\ref{proof2}
\end{IEEEproof}
\subsection{Performance Analysis}
\label{perform}
From the security perspective, we prove that the higher the adversarial ratio is, the higher the probabilities that the adversary can successfully perform attacks on the chain. Similarly, the adversarial ratio also has a negative impact on the performance of the network. In Table~\ref{tab:Compare}, we examine and compare the transaction confirmation time under different adversarial ratio (percentage of stakes in PoS or computational power in PoW that the adversary controls) of a PoW blockchain network (Bitcoin), a PoS network with delayed finality (Cardano), and FedChain's consensus mechanism. The transaction confirmation time of Bitcoin and Cardano is presented in~\cite{Ouroboros}. The transaction confirmation time is the time it takes to reach a CP violation probability $\rm{Pr}_{CP} \leq 0.1 \%$. Based on (\ref{prob}), $\kappa$ can be determined, and then $\kappa$ is multiplied with the time slot duration to calculate the transaction confirmation time. Our time slot duration is set to be 20 seconds (the same as that of Cardano~\cite{Cardanotime}). 

\begin{table}[]
	\small
	\centering
	\caption{Transaction confirmation time in minutes} 
	\begin{tabular}{|>{\raggedright\arraybackslash}m{1.5cm}|>{\raggedright\arraybackslash}m{1.5cm}|>{\raggedright\arraybackslash}m{1.5cm}|>{\raggedright\arraybackslash}m{2cm}|}
		\hline 
		{\centering\arraybackslash}{\textbf{Adversarial Ratio}} &
		{\centering\arraybackslash}{\textbf{Bitcoin }}&
		{\centering\arraybackslash}{\textbf{Cardano}}&
		{\centering\arraybackslash}{\textbf{FedChain's Consensus Mechanism}}\\
		\hline 
		\hline
		0.10             & 50   & 5  &       1           \\ 
		\hline
		0.15             & 80             & 8 & 1.3\\ 
		\hline
		0.20           &     110           & 12         &  1.6                \\ 
		\hline
		0.25         & 150 &  18      &  1.6 \\ 
		\hline
		0.30 &  240                    &  31      & 2  \\
		\hline
		0.35 & 410 & 60  & 2.3  \\
		\hline
		0.40 & 890 & 148  & 2.6  \\
		\hline
		0.45 & 3400 & 663  & 3  \\
		\hline
	\end{tabular}
	\label{tab:Compare}
\end{table}

As observed in Table~\ref{tab:Compare}, the more stakes the adversary controls, the longer the transaction confirmation time is. Moreover, the PVSS protocol no longer ensures unbiased randomness if the adversary controls more than 50\% stakes in a chain. Therefore, it is critical to attract more participants to individual chains in order to increase the network's total stakes and prevent the adversary from controlling more than 50\% of network stakes. In the next section, we will introduce an effective incentive mechanism developed based on a Stackelberg game model that can jointly maximize profits for the participants and significantly enhance the network's performance and security for chain operators.
	\section{Stackelberg Game Formulation}
	\label{Stackel}
	In practice, chains usually announce their block rewards first, and then the stakeholders will decide how much to invest accordingly. Therefore, the interaction between the chains and stakeholders in FedChain can be formulated as a multiple-leaders-multiple-followers Stackelberg game model~\cite{game}. In this game, the leaders are the chains (managed by the chain operators) who first announce their block rewards, and then the stakeholders, i.e., followers, will make their decisions, e.g., how much to invest in each chain.
	\subsection{Stakeholders and Chain Operators}
	FedChain consists of a set $\mathcal{M}$ of $M$ chains and a set $\mathcal{N}$ of $N$ followers. The leaders offers block rewards $\mathbf{R} = (R_1, \ldots, R_M)$. Stakeholders possess stakes with budgets, denoted as $\mathbf{B} = (B_1, \ldots, B_N)$. The stakeholders can use their stakes to take part in the consensus process of every chain to earn additional profits. Particularly, when stakeholder $n$ invests $s_n^m$ to chain $m$, its expected payoff $U^m_n$ is: 
	\begin{equation}
	\label{eq:payoff}
	U^m_n=\dfrac{s_n^m}{s^m_n+\sum_{i \in \mathcal{N}_{-n}} s_i^m}R_m,
	\end{equation}
	where $\mathcal{N}_{-n}$ is the set of all stakeholders except stakeholder $n$. The stakeholders can freely invest within their budgets to any chain, i.e., $\sum_{m=1}^M s_n^m \leq B_n$. Thus, the total payoff of stakeholder $n$ is
	\begin{equation}
	\label{eq:totalpayoff}
	U_n=\sum_{m=1}^{M}U^m_n=\sum_{m=1}^{M}\bigg(\dfrac{s_n^m}{s^m_n+T_m}R_m\bigg),
	\end{equation}
	where $T_m=\sum_{i \in \mathcal{N}_{-n}} s_i^m$ expresses the total stakes invested in chain $m$ by all the other stakeholders.
	\subsection{Game Theoretical Analysis}
	\subsubsection{Followers' strategy}
	To analyze the game, we first examine the existence of the follower sub-game equilibrium in Theorem 3.
	\begin{theorem}
		There exists at least one Nash equilibrium in the follower sub-game.
	\end{theorem}
	\begin{IEEEproof}
		See Appendix~\ref{proof3}.
	\end{IEEEproof}
	Then, we examine the uniqueness of the equilibrium in Theorem 4.
\begin{theorem}
	The follower sub-game equilibrium is unique.
\end{theorem}
\begin{IEEEproof}
		See Appendix~\ref{proof4}.
\end{IEEEproof}
	In this game, the stakeholders can invest any number of stakes within their budgets. However, as shown in Theorem 5, a rational stakeholder will always invest all its budget to maximize its profits regardless other stakeholders' strategies.
	\begin{theorem}
		
		For every follower $n$, the strategies that invest less than its total budget, i.e., $\sum_{m=1}^{M}s^m_n<B_n$, always give lower payoffs than the strategy that invests all the budget, i.e., $\sum_{m=1}^{M}s^m_n=B_n$, regardless of other followers' strategies.
	\end{theorem}
	\begin{IEEEproof}
		See Appendix~\ref{proof5}.
	\end{IEEEproof}
	
	An important result from Theorem 5 is that the strategies which invest less than the total budget can be removed from the strategy space of every follower. Then, we can reformulate the utility function to reflect the budget constraint as follow:
	\begin{equation}
	\begin{split}
	U_n=\sum_{m=1}^{M-1}\bigg(\dfrac{s_n^m}{s^m_n+T_m}R_m\bigg)
	+\dfrac{B_n-\sum_{m=1}^{M-1}s_n^m}{B_n-\sum_{m=1}^{M-1}s_n^m+T_M}R_M.
	\end{split}
	\label{eq:newpay}
	\end{equation}

	With the existence and uniqueness guaranteed, the only question remained is how to find the equilibrium point. Interestingly, for the considered game model, we can prove the exact formula of the equilibrium in Theorem 6.
	\begin{theorem}
		The point where every follower's strategy satisfies $s_n^{*m}=B_n\dfrac{R_m}{\sum_{i=1}^{M}R_i}, \forall m \in \mathcal{M}, \forall n \in \mathcal{N}$ is the unique equilibrium of the follower sub-game.
	\end{theorem}
	\begin{IEEEproof}
		See Appendix~\ref{proof6}
	\end{IEEEproof}
	Then, we can conclude that there is a unique sub-game equilibrium for every fixed leader strategy set, and at the equilibrium the stakeholders will play their optimal strategies, i.e.,
	\begin{equation}
		\label{eq:f_op}
		s_n^{*m}=B_n\dfrac{R_m}{\sum_{i=1}^{M}R_i}, \forall m \in \mathcal{M}, \forall n \in \mathcal{N}.
	\end{equation}
	  
	 This optimal strategy only depends on the stakeholder's total budget and the ratios of block rewards between the chains, i.e., $\sum_{m=1}^{M}s_n^m=\sum_{n=1}^{N}B_n\dfrac{R_m}{\sum_{i=1}^{M}R_i}, \forall m \in \mathcal{M}$. In the next stage, we will analyze the leader strategy to determine the optimal block reward for the leaders.
	\subsubsection{Leader strategy}
\begin{figure}[!]
	\includegraphics[width=\columnwidth]{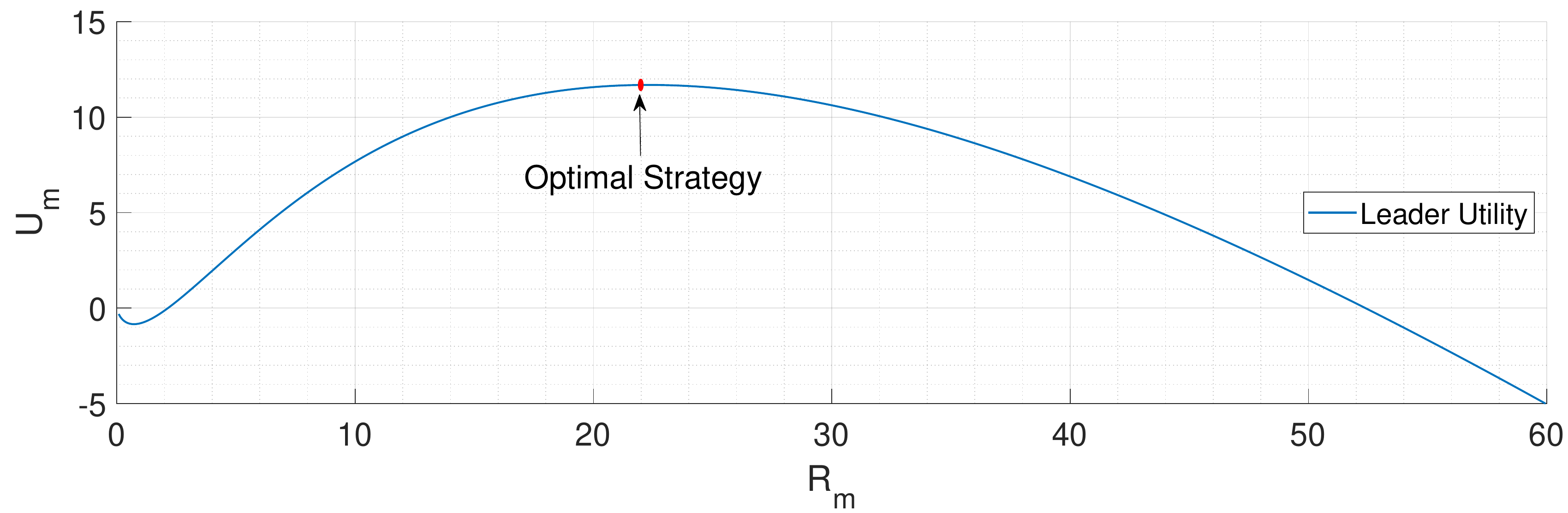}
	\centering
	\caption{An example of the leader's utility function.}
	\label{Fig:leadr_uti}
\end{figure}

	The proposed incentive mechanism for FedChain has two main aims. The first one is to attract stakes to improve the individual chain's performance and security. The second aim is to ensure the decentralization of the system, i.e., encourage the stakeholders to distribute their stakes evenly across all the chains. For these two aims, we propose a utility function $U_m$ for the leaders as follows:
	\begin{equation}
		\label{eq:leader}
		\begin{split}
			U_m&=\sum_{n=1}^N\omega_m^n s_n^{*m}-R_m\\&=
			\sum_{n=1}^N\dfrac{B_nR_m}{\sum_{i=1}^{M}R_i}\lnb{\dfrac{B_nR_m}{\sum_{i=1}^{M}R_i}}-R_m,
		\end{split}
	\end{equation}
	where $\omega_m^n$ is a weight factor which can be defined by $\omega_m^n=\lnb{s^{*m}_n}$. By using the logarithm of the stakes as the weight factor, we can achieve two main aims. In particular, from this designed utility function, a leader can attract more stakes invested to its pool by increasing its block reward. However, at a certain level, if this leader keeps increasing its block reward to get more stakes, its utility will be decreased. As a result, this utility function, as illustrated in Fig.~\ref{Fig:leadr_uti}, encourages the chain operator to set an appropriate level of block reward such that it can attract sufficient stakes to the chain while ensuring that individual stakeholders do not control too much of the network stakes. Moreover, this also discourages the chain operators from setting a too high block reward that will cause the centralization of stakes into a single chain in FedChain. Based on the proposed utility function, we proceed to find the equilibrium of the upper sub-game and the Stackelberg equilibrium of the considered Stackelberg game in Theorem 7.
	\begin{theorem}
		The point where every leader's strategy is $R^*_m=\dfrac{M-1}{M^2}\sum_{n=1}^NB_n\bigg(1+\lnb{\dfrac{B_n}{M}}\bigg)$ and every follower's strategy satisfies $s_n^{*m}=B_n\dfrac{R_m}{\sum_{i=1}^{M}R_i}, \forall m \in \mathcal{M}, \forall n \in \mathcal{N}$ is the unique Stackelberg equilibrium of the considered game.
	\end{theorem}
	\begin{IEEEproof}
		See Appendix~\ref{proof7}.
	\end{IEEEproof}
	Interestingly, the result from Theorem 7 shows that the optimal strategies are the same for all the chain operators. The reason is that since stakes can be transferred, the security of the whole system is as strong as that of the weakest chain. Therefore, the highest utility can only be achieved when every chain is equally secure.  

	\section{Performance Evaluation}
	\label{sim}
	In this section, we conduct experiments and simulations to (i) show that the proposed Stackelberg game can help the stakeholders to maximize their profits, (ii) confirm our analytical results, and (iii) demonstrate that the proposed incentive mechanism can enhance FedChain's security and performance. To this end, we first examine the utility function of a stakeholder to confirm our results from Theorem 6 and show that the Stackelberg game model can help to maximize the stakeholder's profit. After that, to evaluate the security and performance of the FedChain, we implement extensive simulations under various settings. In the simulations, we first show that the rational stakeholders will act according to our proposed Stackelberg game-theoretical analysis. We will then demonstrate that the FedChain's consensus mechanism can satisfy the security properties and attain reasonable performance even under extreme adversarial scenarios. Furthermore, we will show that under the same simulation setting, the proposed dynamic reward scheme achieves better security and performance compared to those of the static reward scheme.  
	\subsection{Simulation Setting}
	First, we examine the utility function of stakeholder 1 in a small case which consists of two stakeholders and three chains. The stakeholders have budgets $\mathbf{B}=[100, 300]$, and the chains set block rewards to be $\mathbf{R}=[10, 20, 30]$. In this experiment, the strategy of stakeholder 2 is fixed according to~\eqref{eq:f_op}. Then, we simulate a system with $N$ stakeholders and $M$ chains under different adversarial models (static and adaptive), reward schemes (static and dynamic), and different adversarial levels (weak, medium, and strong). The simulation parameters are presented in Table~\ref{tab:para}. 
	
	The simulation has several steps as presented in Algorithm 1 (see Appendix~\ref{proof8}). In particular, at the beginning, each stakeholder has a budget $B_i \in [\rm LB, UB]$ generated randomly with uniform distribution. Each chain operator then sets a block reward $R_m$ based on Theorem 7's result in the case of the dynamic reward scheme. In the static reward scheme, $R_m$ are fixed as constants based on several real-world PoS blockchain networks~\cite{Cardano,Algorand2,Tezos}. After the block rewards are set, the stakeholders make their decisions. To find the best strategies for each stakeholder, we employ the Matlab fmincon function~\cite{fmincon}, starting from stakeholder 1. Then, the newly found optimal strategy is fixed for the stakeholder, and the algorithm continues to find the best response for player 2 until stakeholder $N$. After that, the adversary begins to attack. In the static adversary scenario, the adversarial stakes budget $B_A$ is constant and predetermined. In the adaptive adversary scenario, the adversary chooses a number $N_A$ of stakeholders to corrupt, making their stakes to be adversarial stakes, i.e., the adversarial stakes budget is $\sum_{i \in \mathcal{N}_{A}^S} B_i$. Then, we measure the impacts of the adversary on $\rm{Pr}_{CP}$, $\rm{Pr}_{CQ}$, transaction confirmation time, and transaction throughput. Finally, we simulate the stake changes during the epoch by randomly choosing $N_\Delta$ stakeholders and changing their budgets by $\pm \Delta_sB_n, \Delta_s \in (0,1)$. The epoch is then ended, and the simulation moves to the next epoch until the stopping criteria are met, i.e., after $n_e$ epochs.
\begin{table}[!]
	\caption{Parameter setting}
	\label{tab:para}
	\begin{tabular}{|>{\raggedright\arraybackslash}m{1.5cm}|>{\raggedright\arraybackslash}m{1.5cm}|>{\raggedright\arraybackslash}m{1.5cm}|>{\raggedright\arraybackslash}m{1.5cm}|}
		\hline 
		\multicolumn{1}{|>{\centering\arraybackslash}m{1.5cm}|}{\multirow{3}{*}{\textbf{Parameter}}} &
		\multicolumn{1}{>{\centering\arraybackslash}m{1.5cm}|}{\multirow{3}{*}{\textbf{Weak }}}&
		\multicolumn{1}{>{\centering\arraybackslash}m{1.5cm}|}{\multirow{3}{*}{\textbf{Medium}}}&
		\multicolumn{1}{>{\centering\arraybackslash}m{1.5cm}|}{\multirow{3}{*}{\textbf{Strong}}}\\
		&&&\\
		& {\centering\arraybackslash}\textbf{Adversary}& {\centering\arraybackslash}\textbf{Adversary}&{\centering\arraybackslash}\textbf{Adversary} \\
		\hline 
		\hline
		$N$          &  100  &100&100 \\ \hline
		$M$      &      3&3      &    3   \\ \hline
		$\rm LB$           &50&50 &   50    \\ \hline
		$\rm UB$         & 100& 100&  100   \\ \hline
		$\Delta_s$               &  $(0,1)$&  $(0,1)$&  $(0,1)$     \\ \hline
		$B_A$            &    500  & 1000 &1500      \\ \hline
		$N_A$            &     10 &20&30    \\ \hline
		$n_e$                    &   10 &10&10   \\ \hline
	\end{tabular}
\end{table}
	 
	 During the simulation, we measure several important security and performance criteria. First, we measure the stake distribution at the beginning of each epoch to see if the rational stakeholders invest according to our game-theoretical analysis. Then, we examine four different scenarios. In the first two scenarios, we simulate a static adversary who will try to attack the chains under the static and dynamic reward schemes. In the remaining scenarios, an adaptive adversary will try to attack the chains. For each type of adversary, we simulate three different levels of adversary capacity (low, medium, and high) as shown in Table 1. 
	 
	 In terms of security, we measure the CP and CQ violation probabilities. These probabilities can be determined by~\eqref{prob} and~\eqref{prob3}, respectively. In terms of performance, we measure how much the adversaries can negatively impact the transaction confirmation time and transaction throughput. To calculate the transaction confirmation time, for each chain, we find the value of $\kappa$ such that $\rm Pr_{CP}<0.1\%$. For the transaction throughput, we want to examine the case where the adversary wants to reduce the transaction processing capability of one of the chains. Specifically, the adversary will move all its stakes to a chain and participate in the leader selection process. For every block the adversary is elected to be the leader, it creates an empty block without any transaction, thereby reducing the network's transaction throughput. In the simulation, we measure a transaction throughput reduction threshold $\Theta$, such that the probability that the adversary can reduce the transaction throughput more than $\Theta$ is overwhelmingly low (i.e., $\rm Pr_{CQ} <0.1\%$).

 	\subsection{Performance Results}
	\subsubsection{Economical benefits}
\begin{figure}[!]
	\includegraphics[width=.8\columnwidth]{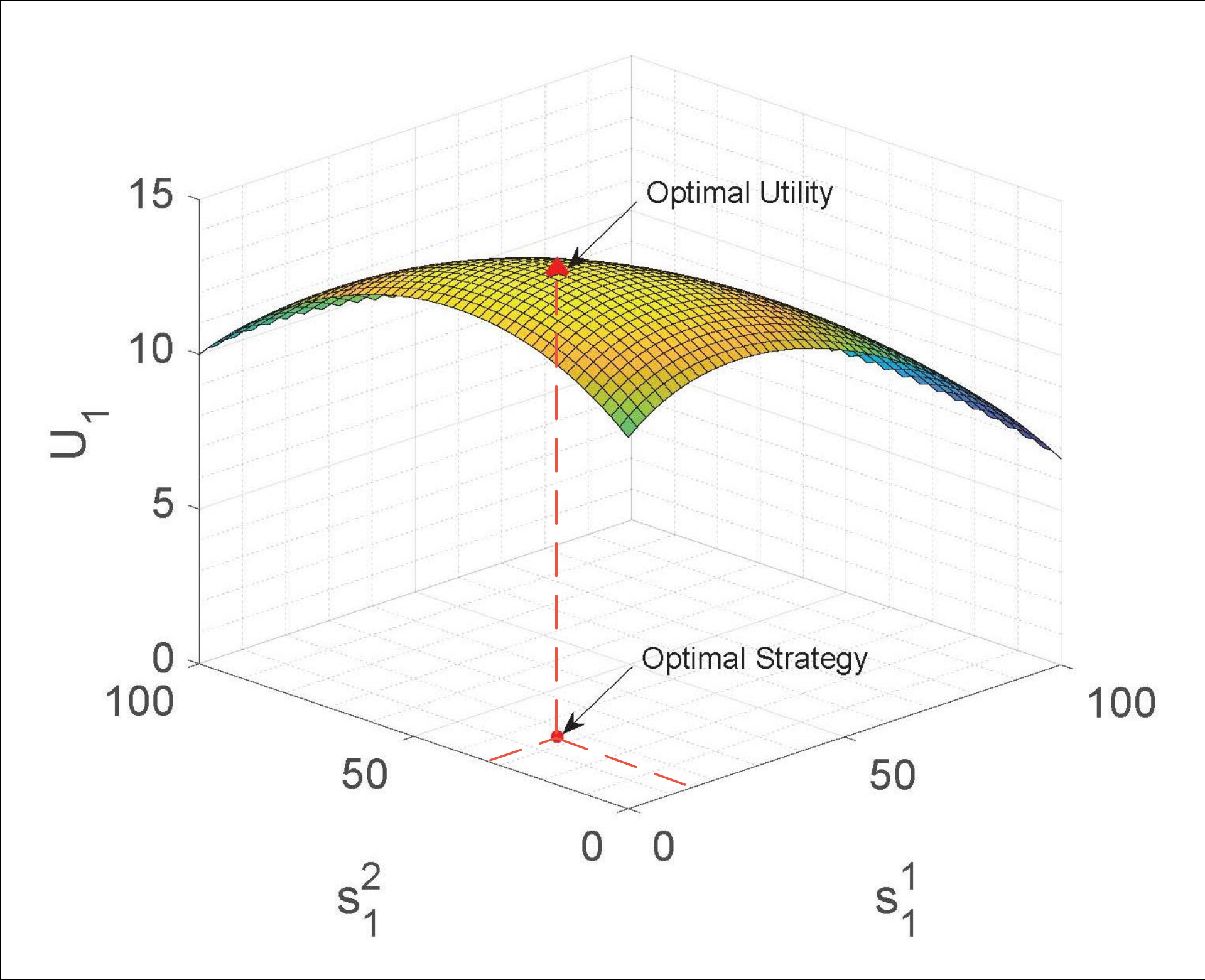}
	\centering
	\caption{Stakeholder's utility function.}
	\label{Fig:Uti_1f}
\end{figure}

 	Fig.~\ref{Fig:Uti_1f} illustrates the utility function of stakeholder 1 in the case where stakeholder 2 invest according to~\eqref{eq:f_op}. As observed from the figure, stakeholder 1 can achieve maximum utility when it also invests according to~\eqref{eq:f_op}. Particularly, stakeholder 1 achieves a utility $U^*_1=15$ with the optimal strategy $\mathbf{s}^{*}_1=[16.6,33.3,50]$. This result shows that our Stackelberg game model can help the stakeholders to achieve maximum profits. Moreover, the ratios between $s^{*1}_1$, $s^{*2}_1$, and $s^{*3}_1$ are the same as the ratios between $R_1$, $R_2$, and $R_3$, which confirms our results in Theorem 6. 
 	\subsubsection{Stake distribution}
\begin{figure}[!]
 	 		\includegraphics[width=\columnwidth]{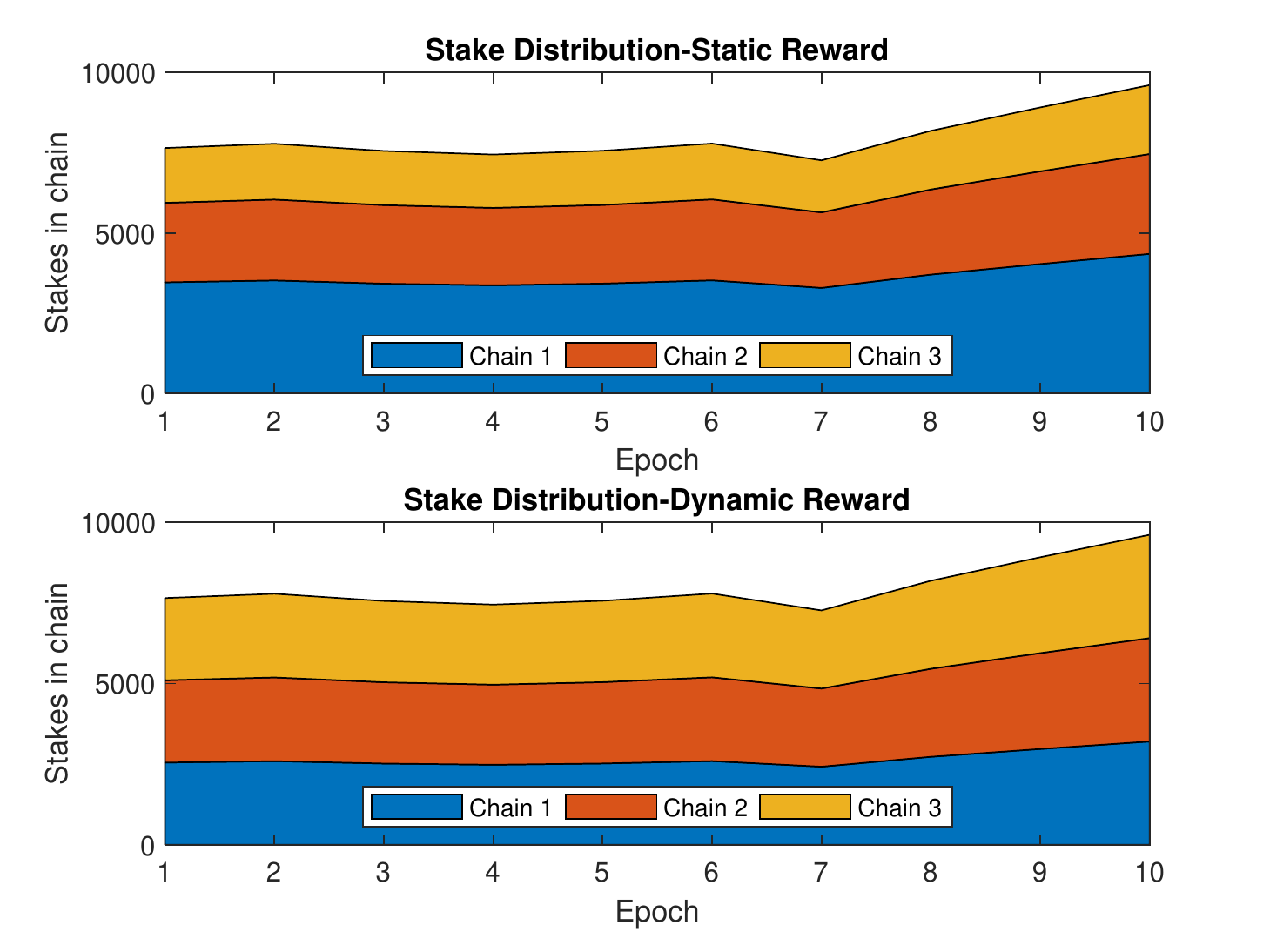}
 	 		\centering
 	 		\caption{Stake distribution.}
 	 		\label{Fig:simu_3c_stakedist}
 	 	\end{figure}

 	 	Fig.~\ref{Fig:simu_3c_stakedist} illustrates the stake distribution at the end of each epoch. As can be seen from the figure, although the total number of stakes vary across the epochs, the ratio of stakes invested in each chain remains unchanged in both the dynamic and static reward schemes. Moreover, we can observe that the stakes are distributed more evenly in the dynamic reward scheme, which is more beneficial to the chains' security and performance. Furthermore, the stake ratios in both schemes equal the ratio of the block rewards, which confirms our analytical results in Theorem 6. 
 	 	\subsubsection{Security properties}
\begin{figure}[!]
		\includegraphics[width=\columnwidth]{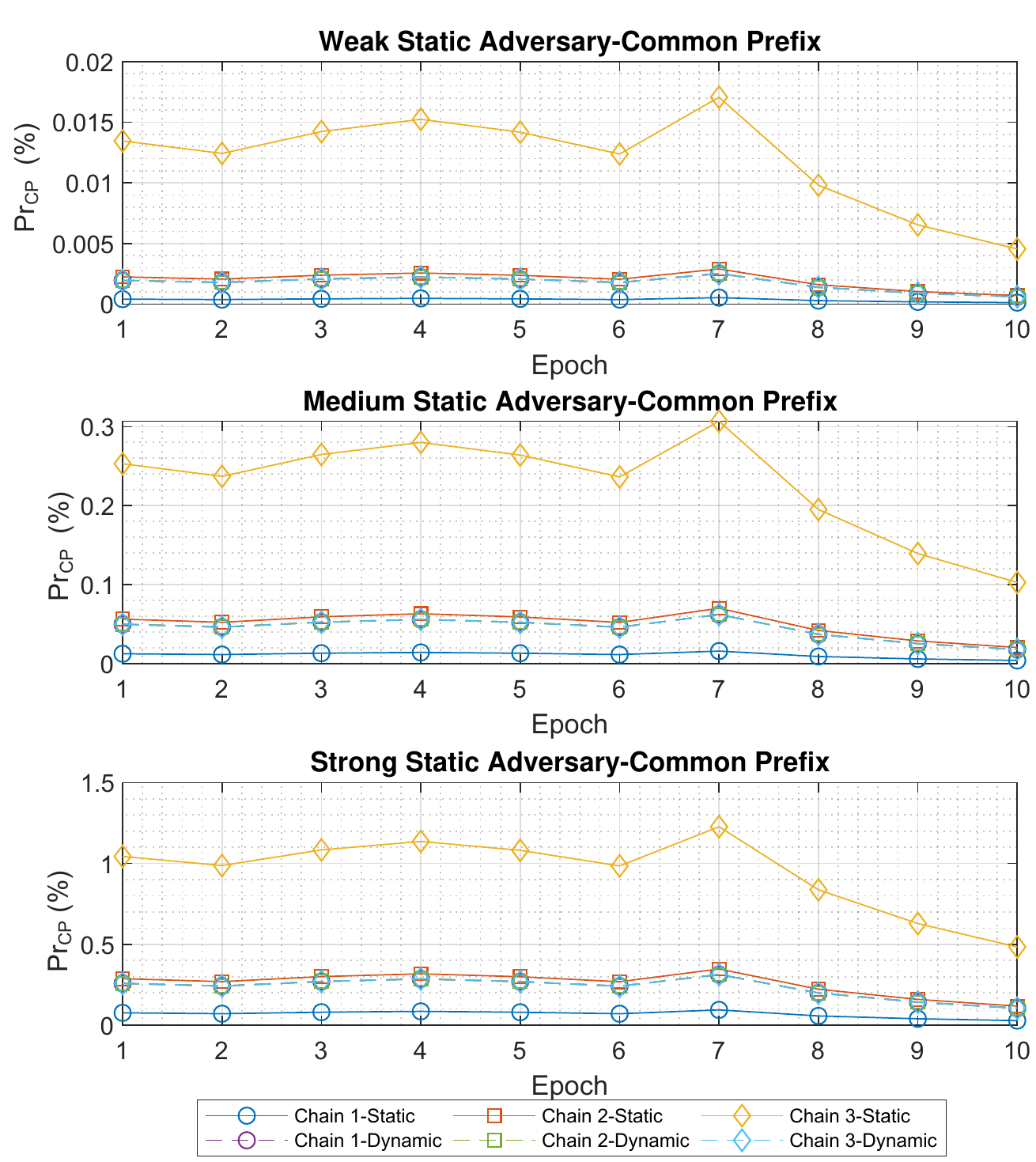}
		\centering
		\caption{$\rm{Pr}_{CP}$ under static adversary settings.}
		\label{Fig:simu_3c_CP_static}
	\end{figure}

\begin{figure}[!]
	\includegraphics[width=\columnwidth]{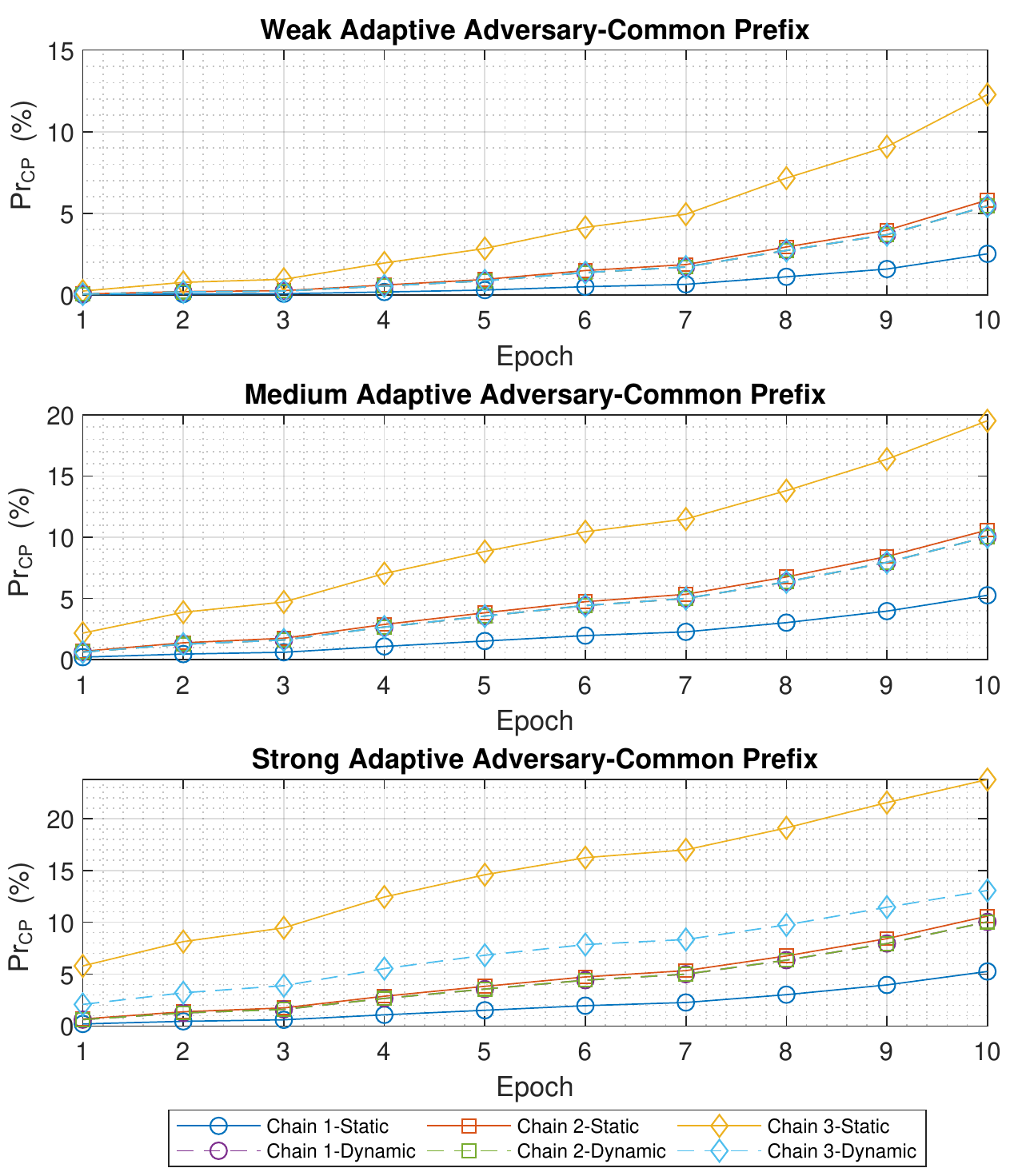}
	\centering
	\caption{$\rm{Pr}_{CP}$ under adaptive adversary settings.}
	\label{Fig:simu_3c_CP_adapt}
\end{figure}

 	 	Fig.~\ref{Fig:simu_3c_CP_static} and Fig.~\ref{Fig:simu_3c_CP_adapt} illustrate $\rm{Pr}_{CP}$ of each chain at the end of each epoch under the static and adaptive adversary settings, respectively. From the figures, we can observe that the more stakes the adversary controls, the higher chance it can violate the security of the system. For example, in the static adversary setting, with a low budget (weak adversary), $\rm{Pr}_{CP}$ is at most 0.02\%, whereas this probability increases to 1.5\% in case of an adversary with a high budget (strong adversary). Secondly, the total system stakes have different effects on the chains' security under the static and adaptive adversary setting. For instance, the system has the highest stakes in the last epoch. At this epoch, $\rm{Pr}_{CP}$ achieve the lowest value under the static adversary because the static adversary has a fixed budget. However, $\rm{Pr}_{CP}$ achieve the highest value under the adaptive adversary setting because the adaptive adversary can corrupt the stakeholders with the most stakes. Therefore, it is crucial to not only attract more stakes to the system but also to incentivize more diversity, i.e., encourage the stakeholders to split their stakes across more chains. We can observe the effect of such diversity between the dynamic and the static reward schemes. Although the total network stakes are the same, the dynamic scheme, which encourages equal stakes distribution, achieves much lower $\rm{Pr}_{CP}$, e.g., at most 14\% compared to 24\% of the static reward scheme.

\begin{figure}[!]
	\includegraphics[width=\columnwidth]{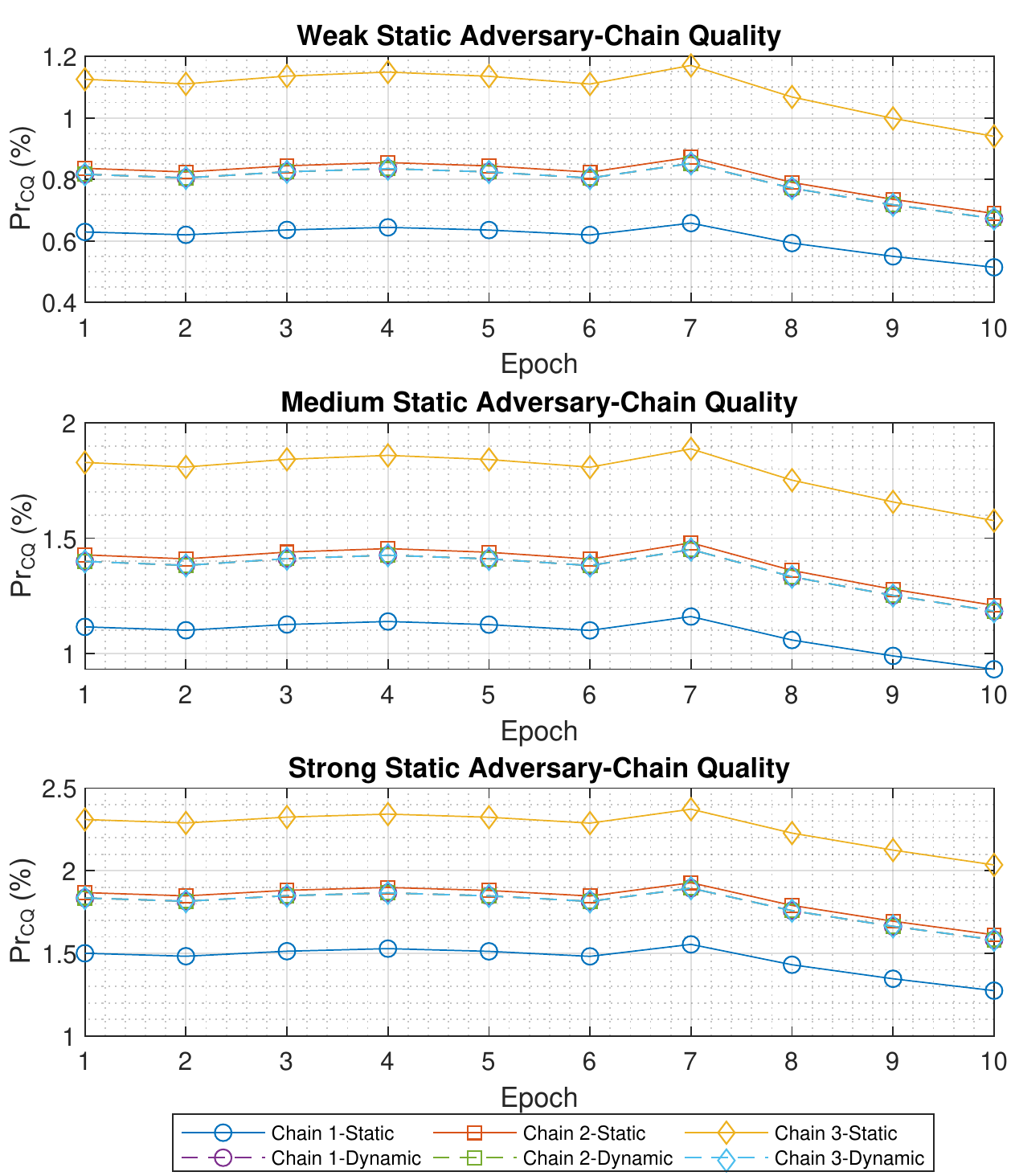}
	\centering
	\caption{$\rm{Pr}_{CQ}$ under static adversary settings.}
	\label{Fig:simu_3c_CQ_static}
\end{figure}

\begin{figure}[!]
	\includegraphics[width=\columnwidth]{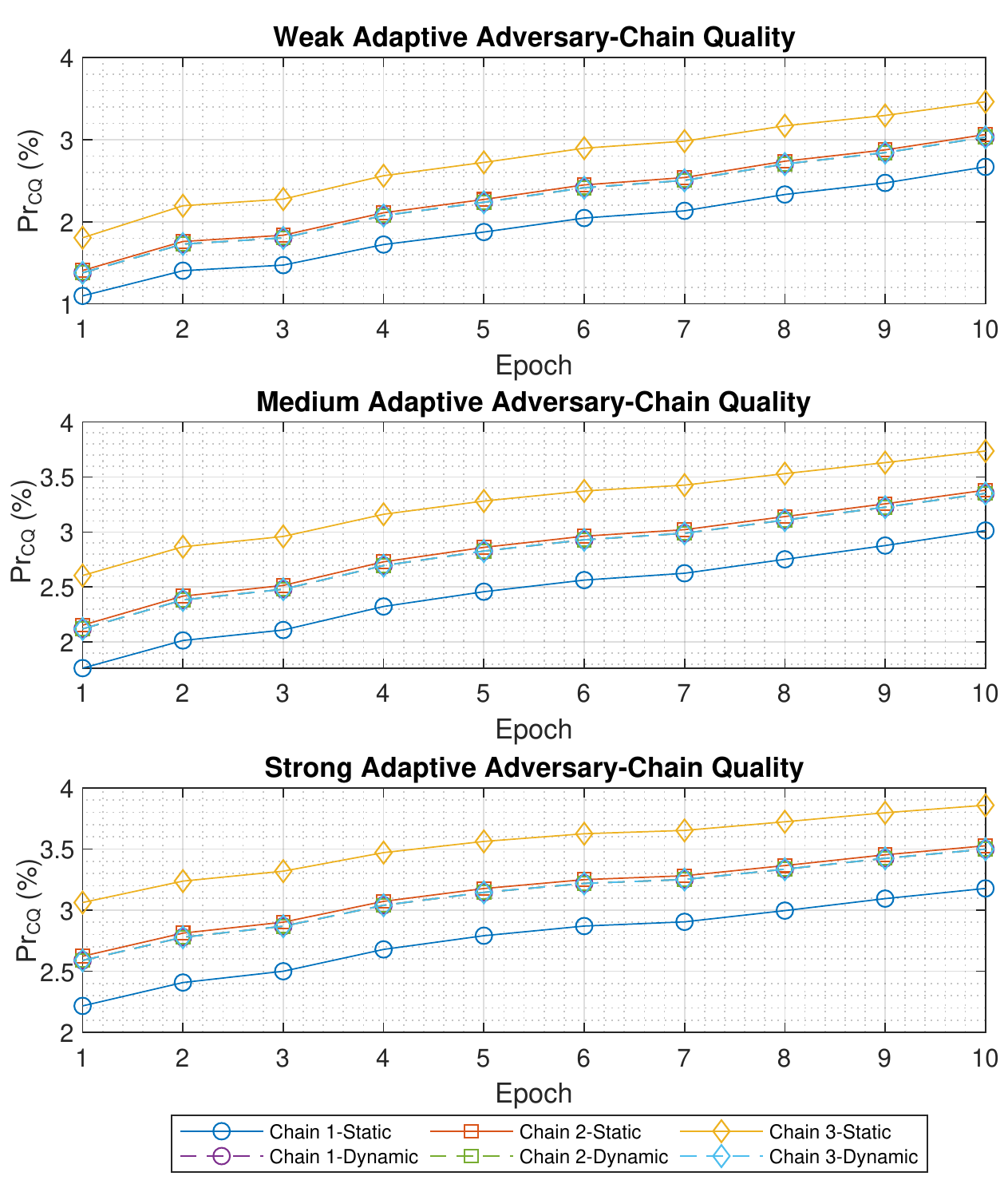}
	\centering
	\caption{$\rm{Pr}_{CQ}$ under adaptive adversary settings.}
	\label{Fig:simu_3c_CQ_adapt}
\end{figure}

Fig.~\ref{Fig:simu_3c_CQ_static} and Fig.~\ref{Fig:simu_3c_CQ_adapt} illustrate $\rm{Pr}_{CQ}$ of each chain under the static and adaptive adversary settings, respectively. Similar to the $\rm{Pr}_{CP}$, we can draw several conclusions from examining $\rm{Pr}_{CQ}$. Firstly, the stronger the adversary is, the higher chance it violates system security. For example, in the weak adaptive adversary scenario, $\rm{Pr}_{CQ}$ is at most 1.2\%, whereas this probability increases to 2.4\% in the case of a strong adaptive adversary. Generally, $\rm{Pr}_{CQ}$ gets higher in the case of the adaptive adversary. The reason is that according to the simulation setting, the adversary can corrupt more stakes compared to $B_A$ in the case of the static adversary. Secondly, similar to the results of $\rm{Pr}_{CP}$, $\rm{Pr}_{CQ}$ is inversely proportional to the total system stakes in the case of the static adversary, and it is proportional to the total system stakes in the case of the adaptive adversary. As a result, we can observe that the dynamic scheme achieves lower $\rm{Pr}_{CQ}$, e.g., at most 14\% $\rm{Pr}_{CQ}$ compared to 24\%. Moreover, since the security of the system is only as good as that of its weakest chain (especially with the SPV proof mechanism), it can be observed that the dynamic reward scheme achieves better security compared to the static reward scheme, i.e., the chains of the dynamic reward scheme always achieve better $\rm{Pr}_{CP}$ and $\rm{Pr}_{CQ}$ compared to those of the weakest chain under the static reward scheme (i.e., Chain 3).


\subsubsection{Performance properties}
\begin{figure}[!]
	\includegraphics[width=\columnwidth]{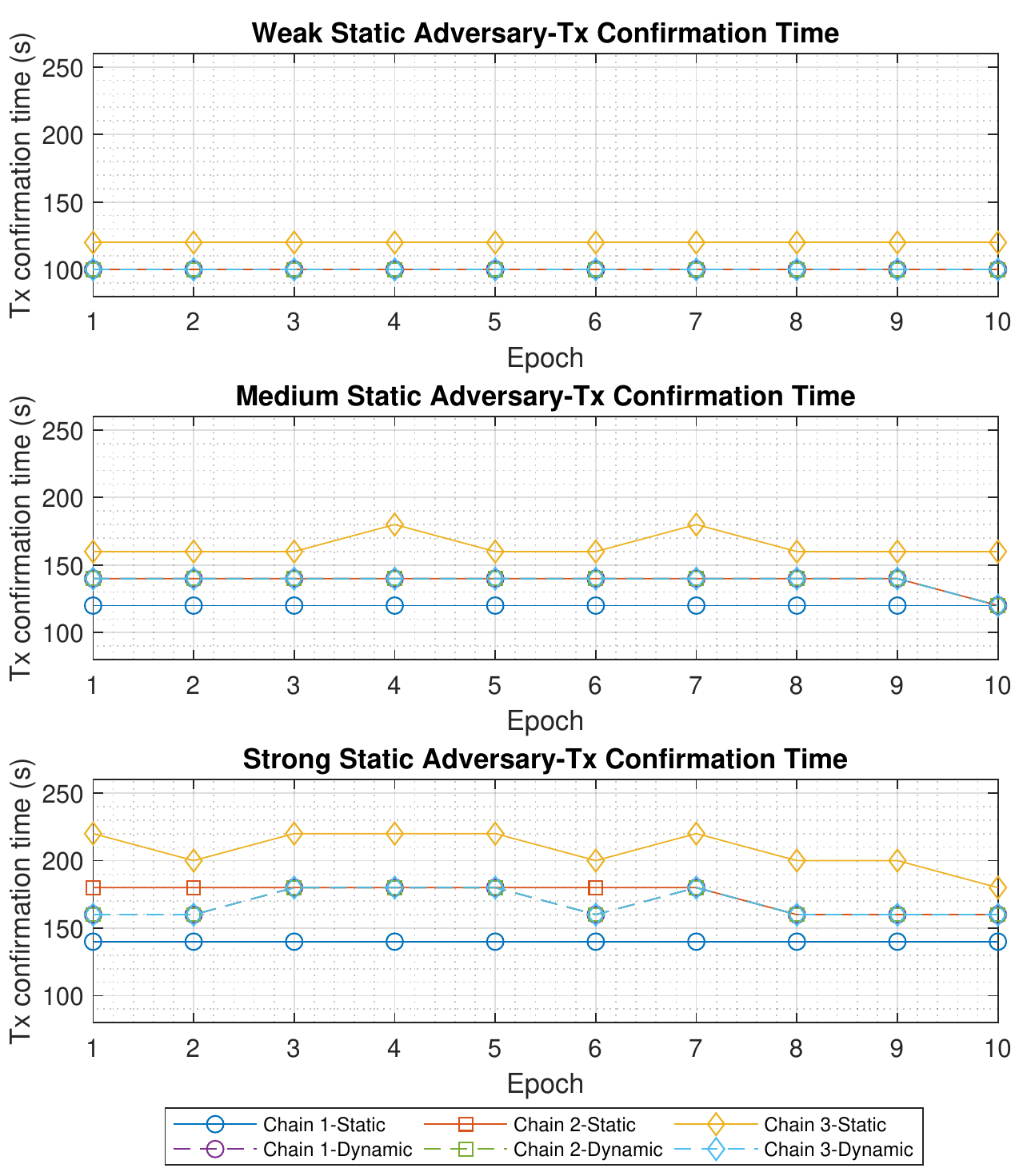}
	\centering
	\caption{Transaction confirmation time under static adversary settings.}
	\label{Fig:simu_3c_Tx_static}
\end{figure}

\begin{figure}[!]
	\includegraphics[width=\columnwidth]{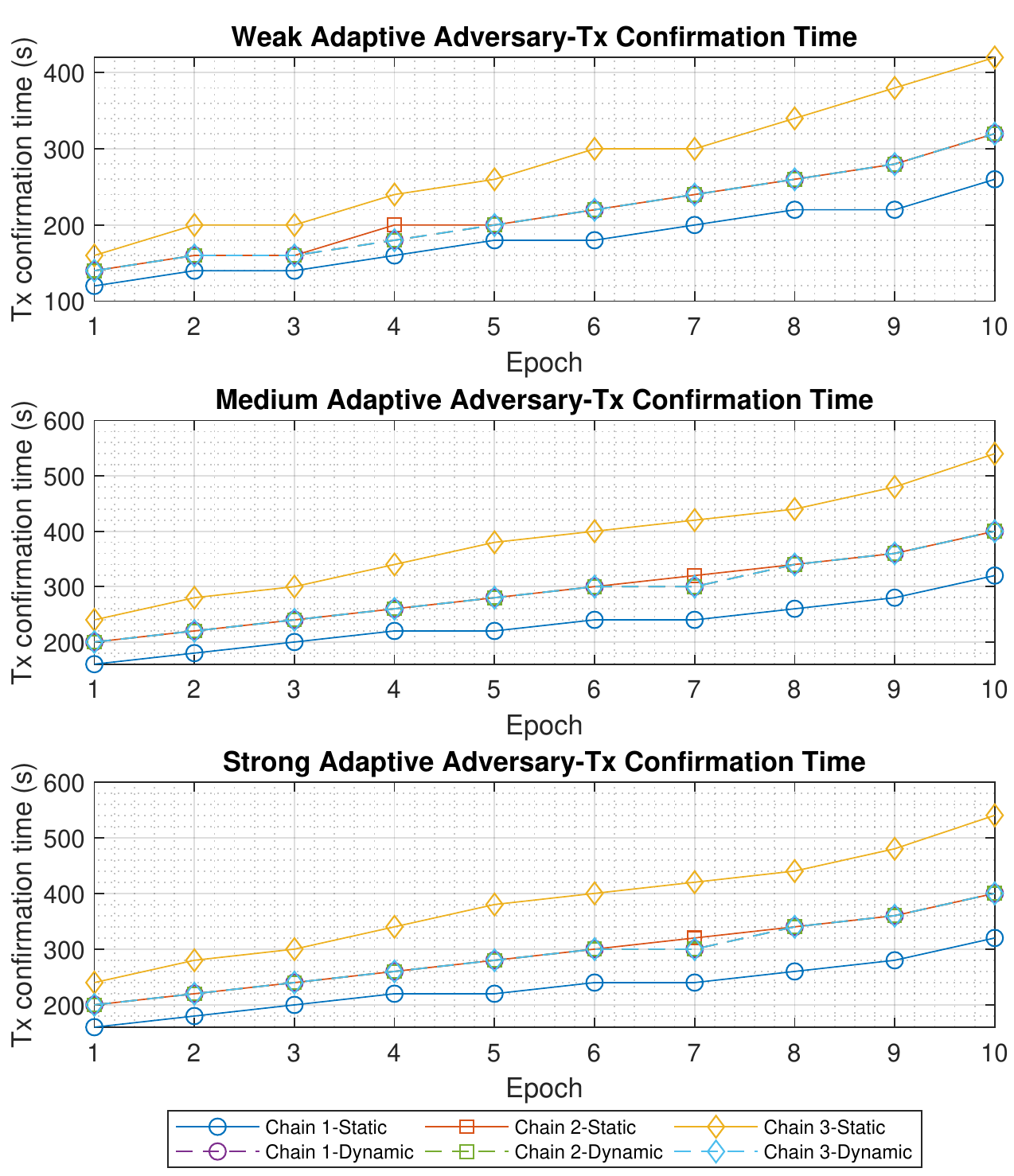}
	\centering
	\caption{Transaction confirmation time under adaptive adversary settings.}
	\label{Fig:simu_3c_Tx_adapt}
\end{figure}

		Fig.~\ref{Fig:simu_3c_Tx_static} and Fig.~\ref{Fig:simu_3c_Tx_adapt} illustrate the transaction confirmation time of each chain under the static and adaptive adversary settings, respectively. From the figures, we can observe that the stronger the adversary is, the more it can negatively affect the system performance. For example, the chains takes at most 120 seconds to confirm a transaction in case of a weak static adversary, but it takes up to 220 seconds in case of a strong static adversary. This is because the transaction confirmation time is directly related to $\rm{Pr}_{CP}$. A stronger adversary has a higher chance to violate the CP property, and thus the users have to wait longer to confirm a transaction. Moreover, we can also observe that the transaction confirmation time is inversely proportional to the total stakes of the system in the static adversary settings, whereas the opposite holds true in the adaptive adversary settings. The reason is the same as that of the $\rm{Pr}_{CP}$ scenarios, i.e., the adaptive adversary can corrupt more stakes, whereas $B_A$ of the static adversary is fixed. Furthermore, the transaction confirmation time of the three chains under the dynamic reward schemes is always better than at least two chains under the static reward scheme. 

\begin{figure}[!]
		\includegraphics[width=\columnwidth]{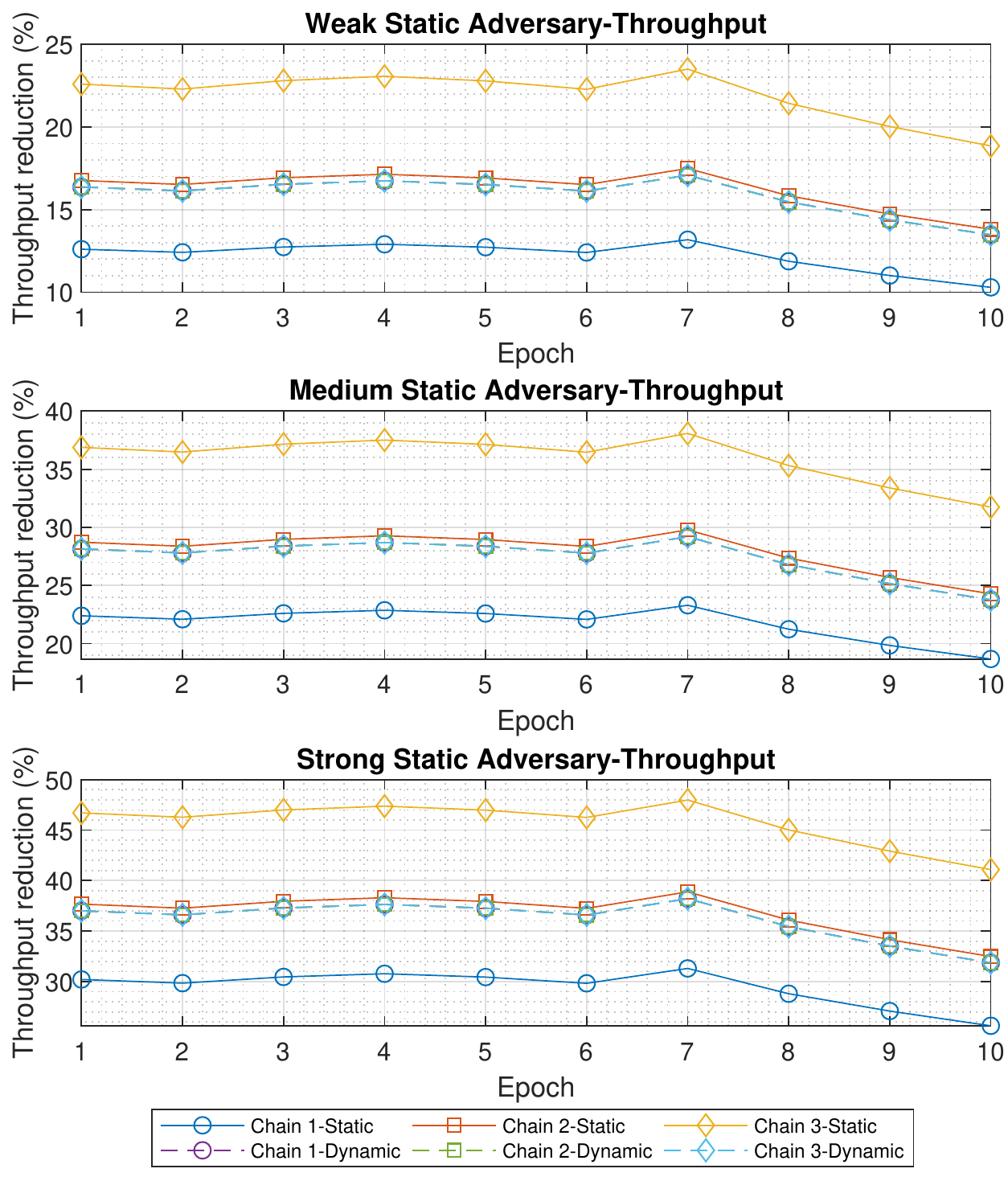}
		\centering
		\caption{Transaction throughput reduction under static adversary settings.}
		\label{Fig:simu_3c_Tp_static}
	\end{figure}

\begin{figure}[!]
				\includegraphics[width=\columnwidth]{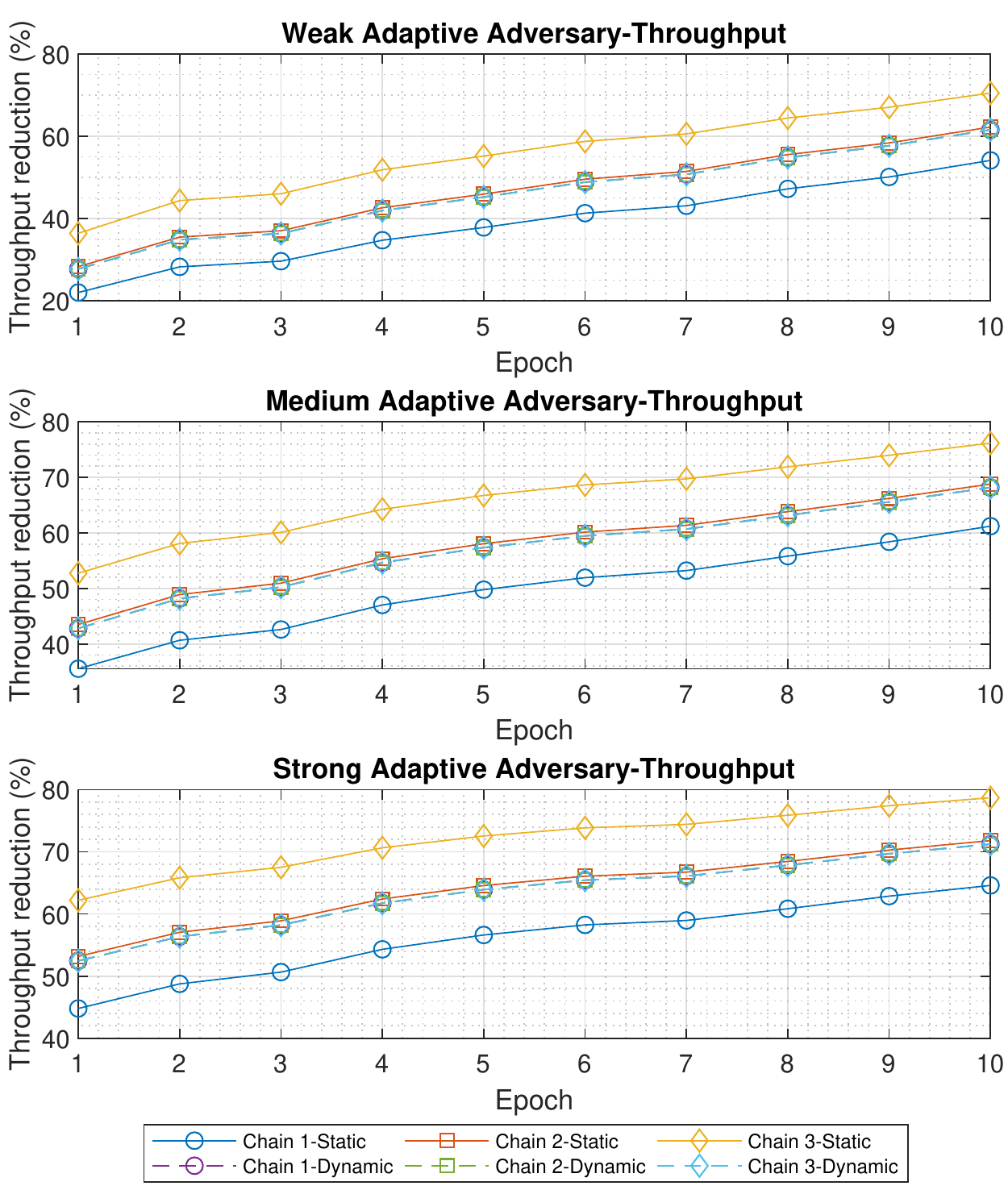}
				\centering
				\caption{Transaction throughput reduction under adaptive adversary settings.}
				\label{Fig:simu_3c_Tp_adapt}
			\end{figure}

	Fig.~\ref{Fig:simu_3c_Tp_static} and Fig.~\ref{Fig:simu_3c_Tp_adapt} illustrate the transaction throughput reduction percentages of each chain under the static and adaptive adversary setting, respectively. Similar to the previous scenarios, we can observe that a stronger adversary can cause more negative impacts on the system performance, e.g., a weak static adversary can reduce the throughput by at most 24\%, whereas the strong static adversary can reduce the throughput by nearly 50\%. Moreover, it can be observed that as the system has more stakes, the static adversary becomes weaker, whereas the adaptive adversary becomes stronger, similar to the previous scenario. Finally, one can observe that the dynamic reward scheme can achieve a better overall performance compared to that of the static reward scheme (the performances of the three chains in the dynamic scheme are better than those of at least two chains in the static scheme). 
	
	\section{Conclusion}
	\label{sec:Sum}
	In this paper, we have introduced FedChain, an effective framework for federated-blockchain systems together with a cross-chain transfer protocol to facilitate the secure and decentralized transfer of tokens between the blockchains. In this framework, we have proposed a novel consensus mechanism which can satisfy the CP, CG, and CQ properties, prevent various blockchain-specific attacks, and achieve better transaction confirmation time compared to existing consensus mechanisms. Robust theoretical analyses have been then conducted to prove FedChain's consensus mechanism security and performance properties. After that, a Stackelberg game model has been developed to examine the interactions between the stakeholders and the blockchains managed by chain operators. This model can provide additional profits for the stakeholders and enhance the security and performance of the blockchains. Through analyses of the Stackelberg game model, we can prove the uniqueness of the Stackelberg equilibrium and find the exact formula for this equilibrium. These results are especially important for the stakeholders to determine their best investment strategies and for the chain operators to design the optimal policy, i.e., block rewards. Finally, extensive experiments and simulations have been conducted to show that our proposed framework can help stakeholders to maximize their profits and the chain operator to design appropriate parameters to enhance FedChain's security and performance.
	
	%
	\bibliographystyle{IEEE}

\cleardoublepage
	\appendices
%
	
	\section{Proof of Theorem 1}
\label{proof1}
We first prove that our FedChain's consensus mechanism can satisfy the CP property in Lemma 1.
\begin{lemma}
	The probability that FedChain's consensus mechanism violates the common prefix property with parameter $\kappa \in \mathbb{N}$ is less than or equal to $(1-\gamma)^\kappa$. 
\end{lemma}
\begin{IEEEproof}
In order to violate the CP property, the adversary must have two forks with at least $\kappa$ conflicting blocks, and both forks must be accepted by the honest stakeholders. However, an honest stakeholder will accept only one fork in the same time slot. Therefore, the adversary must (i) create a fork, (ii) have the honest stakeholders accept it, (iii) create another fork with a conflicting block at a later time slot, and (iv) have the honest stakeholders accept the new fork. We will prove that the adversary can only do that if it is elected to be the leader for $\kappa$ consecutive blocks.

Without loss of generality, assume that the adversary is elected to be the leader at time slot $\rm s^l_1$, $\rm s^l_2$, and $\rm s^l_4$. This means that at $\rm s^l_3$ an honest stakeholder is elected to be the leader. Assume that the adversary wants to create two conflicting forks $\mathcal{C}_1,\mathcal{C}_2$. Let $\mathcal{B}^{i}_j$ denote the block from fork $\mathcal{C}_i$ at time slot $\rm s^l_j$. Firstly, at $\rm s^l_1$ and $\rm s^l_2$, the adversary broadcasts blocks $\mathcal{B}^{1}_1$ and $\mathcal{B}^{1}_2$. At this point, the adversary can create fork $\mathcal{C}_2$ with different blocks, i.e., $\mathcal{B}^{2}_1 \ne \mathcal{B}^{1}_1$ and $\mathcal{B}^{2}_2 \ne \mathcal{B}^{1}_2$, and has both forks accepted by the honest stakeholder (some honest stakeholders will adopt $\mathcal{C}_1$ while some will adopt $\mathcal{C}_2$). 

However, at $\rm s^l_3$, the honest leader will either choose one of the two forks to adopt. Assume that the leader chooses $\mathcal{C}_1$, it will add block $\mathcal{B}^{1}_3$ to the chain, and the fork $\mathcal{C}_2$ will be discarded by all honest stakeholders. Next, at $\rm s^l_4$, the adversary is elected to be the leader and can create a fork again. At this point, the adversary can try to broadcast $\mathcal{C}_2$ to the honest stakeholder again with block $\mathcal{B}^{2}_1 \ne \mathcal{B}^{1}_1$ (e.g., to gain back the tokens spent in block $\mathcal{B}^{1}_1$). Nevertheless, any change in a block's content results in a different block's hash, and the block's hash is linked to its previous block. Thus, $\mathcal{B}^{1}_1$ cannot be changed unless block $\mathcal{B}^{1}_3$ is changed. However, since the leader of $\mathcal{B}^{1}_3$ is honest, the block will not be changed (due to consensus rules $I_2$ and $I_3$).

Moreover, the leader election is conducted at the beginning of each epoch, and as long as the PVSS protocol is secure (proven in~\cite{PVSS}), any honest stakeholder can obtain and verify the correct leader list if that stakeholder is online at least once during the epoch (thanks to consensus rule $I_1$). Since the epoch is long (e.g., 5 days~\cite{Ouroboros}), we can assume that every honest stakeholder will have the correct leader list. Therefore, the adversary also cannot broadcast $\mathcal{B}^{1}_3 \ne \mathcal{B}^{2}_3$ on its own since it is not the designated leader. Thus, the adversary must include $\mathcal{B}^{1}_3$ and every block before that. Otherwise, it will create an invalid fork that will be rejected due to due to consensus rule $I_4$. 

As a result, the part of the chain from the first block to the latest honest block (e.g., until $\mathcal{B}^1_3$ in the above analysis) is confirmed by every honest user. Therefore, the adversary can only create forks with $\kappa$ last blocks different from the honest fork if it is elected to be the leader for $\kappa$ consecutive blocks. Since $(1-\gamma)$ is the ratio of adversarial stakes in the total network stakes, the probability that the adversary is elected to be the leader for $\kappa$ consecutive blocks is 
\begin{equation}
	\label{prob}
	\rm{Pr}_{CP}=(1-\gamma)^\kappa,
\end{equation}
which is also the probability that the CP property is violated.
\end{IEEEproof}

Then, we prove that our FedChain's consensus mechanism satisfies the CG property in Lemma 2.
\begin{lemma}
	FedChain's consensus mechanism satisfies the chain growth property 
\end{lemma}
\begin{IEEEproof}
	Even if a new block is not broadcast during a time slot, an empty block will be added to the chain. Therefore, the CG property will always be satisfied.  
\end{IEEEproof}
Next, we prove that FedChain's consensus mechanism can satisfy the CQ property in Lemma 3.
\begin{lemma}
	The probability that FedChain's consensus mechanism violates the ideal chain quality property with parameters $l, \mu$ over $l$ blocks is no more than $1-\exp\bigg(\dfrac{l(\gamma-1) \delta^2}{2}\bigg)$.
\end{lemma}
\begin{IEEEproof}
	We can characterize the block adding process among the honest stakeholders and the adversary as a binomial random walk~\cite{stat}. During the considered $l$ slots, the leader election processes can be considered independent Bernoulli trials $X_1, \ldots, X_l$ such that, for $1 \le i \le l$, ${\rm Pr} [X_i=0]=\gamma$ and ${\rm Pr} [X_i=1]=1-\gamma$. Then, the expected value of the trials is $\mathbb{E}[X]=\sum_{i=1}^l (1-\gamma)$. Applying the Chernoff bound~\cite{stat}, the probability that the adversary creates less than $l(1-\mu)$ blocks, i.e., $X=\sum_{i=1}^l X_i < l(1-\mu)$, is
	\begin{equation}
		\label{prob2}
		\begin{split}
			{\rm Pr}[X< (1-\delta)\mathbb{E}[X]]&= \exp\bigg(\dfrac{-\mathbb{E}[X] \delta^2}{2}\bigg),\\
			{\rm Pr}[X< (1-\delta)l(1-\mu)]&=\exp\bigg(\dfrac{(\mu-1)l \delta^2}{2}\bigg),
		\end{split}
	\end{equation}
	where $\delta$ is any real number such that $0<\delta \le 1$. In the case of \textit{ideal CQ}~\cite{sec}, we have $\mu=\gamma$, and the ideal CQ violation probability is
	\begin{equation}
		\label{prob3}
		\begin{split}
			{\rm Pr_{CQ}}=1-\exp\bigg(\dfrac{l(\gamma-1) \delta^2}{2}\bigg)
		\end{split}
	\end{equation}
\end{IEEEproof}

From Lemma 1, we have the CP violation probability $\rm{Pr}_{CP}=(1-\gamma)^\kappa$ which decreases exponentially as $\kappa$ grows. Then, we proved in Lemma 2 that the CG property will always be satisfied. Finally, from Lemma 3, we have the CQ violation probability ${\rm Pr_{CQ}}=1-\exp\bigg(\dfrac{l(\gamma-1) \delta^2}{2}\bigg)$ which decreases exponentially as $l$ and $\delta$ grow. Thus, all the three violation probabilities can be satisfied with overwhelming probabilities, and the proof is completed.
	\section{Proof of Theorem 2}
\label{proof2}
We prove FedChain's consensus mechanism's ability to prevent each attack as follows:
\begin{itemize}
	\item{\textbf{Double-spending attack}} To double-spend, the attacker has to either create a conflicting transaction in the same fork, i.e., create $\rm Tx1$ in $\mathcal{B}^1_i$ and $\rm Tx2$ in $\mathcal{B}^1_j$, or create two transactions in two forks, i.e., create $\rm Tx1$ in $\mathcal{B}^1_i\in \mathcal{C}_1$ and $\rm Tx2$ in $\mathcal{B}^2_j\in \mathcal{C}_2$. For the first approach, $\rm Tx2$ is not a valid transaction and will be rejected. For the second approach, if the CP property is not violated, only one of $\mathcal{C}_1$ and $\mathcal{C}_2$ will be confirmed. Thus, this attack is prevented.
	
	\item{\textbf{Grinding attack:}} The seeds for leader and committee selection are created by the committee via the PVSS protocol in FedChain's consensus mechanism. Therefore, grinding attacks are prevented.
	
	\item{\textbf{Nothing-at-stake attacks:}} Although the adversary can create valid forks, they are not confirmed. Therefore, the vendors only have to wait until a transaction is confirmed. Thus, this attack is prevented as long as the CP property holds. 
	
	\item{\textbf{Bribe attacks:}} In the considered adaptive adversary model, the adversary can only corrupt honest stakeholders with a delay. Since the adversary cannot know who is the leader in advance, the adversary cannot bribe the leaders. Thus, this attack is prevented.
	
	\item{\textbf{Transaction denial attack:}} With the liveness property, a transaction will eventually be included in a block created by an honest leader. As a result, this attack can be prevented as long as the liveness property (or the CG and CQ properties) holds.
	
	\item{\textbf{Long-range attack:}} FedChain's consensus mechanism can prevent this attack by locking committee members' stakes during their designated epochs. 
\end{itemize} 
	\section{Proof of Theorem 3}
\label{proof3}
Let $\mathcal{S}_n$ denote the strategy space of follower $n$. Then, any strategy $\mathbf{s}_n=[s^1_n, \ldots, s^M_n]$ that satisfies 
	\begin{equation}
	\label{eq:strat}
	\sum_{m=1}^M s^m_n \leq B_n,
\end{equation}
is a feasible strategy of follower $n$, i.e., $\mathbf{s}_n \in \mathcal{S}_n$. We first prove $\mathcal{S}_n$ to be compact and convex $\forall n \in \mathcal{N}$ in Lemma 4.
\begin{lemma}
	$\mathcal{S}_n$ is compact and convex $\forall n \in \mathcal{N}$.
\end{lemma}
\begin{IEEEproof}
	Let $\mathbf{s}_n$ and $\mathbf{s'}_n$ be any two different strategies in $\mathcal{S}_n$. To prove $\mathcal{S}_n$ is convex, we prove that any convex combination of $\mathbf{s}_n$ and $\mathbf{s'}_n$ is in $\mathcal{S}_n$, i.e.,
	\begin{equation}
		\label{eq:set}
		\begin{split}
			&\lambda \mathbf{s}_n + (1-\lambda)\mathbf{s'}_n=[\lambda s^1_n+(1-\lambda)s^{'1}_n,\ldots,\lambda s^M_n+\\&(1-\lambda)s^{'M}_n] \in \mathcal{S}_n, \forall \lambda \in (0,1), \forall \mathbf{s}_n, \mathbf{s'}_n \in \mathcal{S}_n.
		\end{split}
	\end{equation}
	Since $\lambda s^m_n+(1-\lambda)s^{'m}_n \leq max\{s^m_n, s^{'m}_n\}, \forall m \in \mathcal{M}$, we have
	\begin{equation}
		\label{eq:set2}
		\sum_{m=1}^M \bigg(\lambda s^1_n+(1-\lambda)s^{'1}_n\bigg) \leq max\bigg\{\sum_{m=1}^M s^m_n, \sum_{m=1}^M s^{'m}_n\bigg\} \leq B_n.
	\end{equation}
	From~\eqref{eq:set2}, all convex combinations of $\mathbf{s}_n$ and $\mathbf{s'}_n$ satisfy~\eqref{eq:strat}, and thus they all lie in $\mathcal{S}_n$. As a result, $\mathcal{S}_n$ is convex. Moreover, since $\mathcal{S}_n$ is closed and bounded, it is compact.
\end{IEEEproof}
Then, we prove that $U_n$ is concave in Lemma 5.
\begin{lemma}
	$U_n$ is concave over $S_n$, $\forall n \in \mathcal{N}$.
\end{lemma}
\begin{IEEEproof}
	We have:
	\begin{equation}
		\label{eq:um}
		\dfrac{\partial^2 U^m_n}{\partial (s^m_n)^2}=\dfrac{-2R_mT_m}{(s^m_n+T_m)^3} \le 0.
	\end{equation}
	Thus, $U_n^m$ is concave over $\mathcal{S}_n$. Then, $U_n= \sum_{m=1}^{M} U_n^m$ is also concave over $\mathcal{S}_n$.
\end{IEEEproof}
According to~\cite{game}, if $S_n$ is compact and convex and $U_n$ is quasi-concave $\forall n \in \mathcal{N}$, there exists at least one Nash equilibrium. It follows from Lemma 4 and 5 that the follower sub-game satisfies these conditions, and thus the proof of this Theorem is complete.
	\section{Proof of Theorem 4}
\label{proof4}
According to Rosen's theorem~\cite{Rosen}, a sufficient condition to guarantee the uniqueness of the equilibrium is that the matrix $[\textbf{G}(\mathbf{s},\omega)+\textbf{G}^{T}(\mathbf{s},\omega)]$ is negative definite for a fixed $\omega > 0$. $\textbf{G}(\mathbf{s},\omega)$ can be calculated by:

\begin{equation}
	\small
	\textbf{G}(\mathbf{s},\omega)=
	\begingroup 
	\begin{bmatrix}
		\omega_1\dfrac{\partial^2 U_1}{\partial s^1_1\partial s^1_1} &\omega_1\dfrac{\partial^2 U_1}{\partial s^2_1\partial s^1_1}
		&\cdots
		&\omega_1\dfrac{\partial^2 U_1}{\partial s^M_1\partial s^1_N}\\[1em]
		\omega_1\dfrac{\partial^2 U_1}{\partial s^1_1\partial s^2_1}      
		&\omega_1\dfrac{\partial^2 U_1}{\partial s^2_1\partial s^2_1}
		&\cdots
		&\omega_1\dfrac{\partial^2 U_1}{\partial s^M_1\partial s^2_N}
		\\[1em]		 	
		\vdots      
		&\vdots
		&\ddots
		&\vdots
		\\[1em]		 	
		\omega_N\dfrac{\partial^2 U_N}{\partial s^1_N\partial s^M_1}      &\omega_N\dfrac{\partial^2 U_N}{\partial s^2_N\partial s^M_1}
		&\cdots
		&\omega_N\dfrac{\partial^2 U_N}{\partial s^M_N\partial s^M_N}    
	\end{bmatrix}
	\endgroup 
	\label{eq:t1}
\end{equation}

Let $\omega_n=1,\forall n \in \mathcal{N}$, $\textbf{G}(\mathbf{s},\omega)$ can be rewritten as~\eqref{eq:t2}.
\begin{figure*}[!]
	\begin{equation}
		\begingroup 
		\begin{bmatrix}
			\Phi_1^{1} &0\cdots0&0&&\cdots&&\phi_1^{1}&0\cdots0&0&&\cdots&&\phi_1^{1}&0\cdots0&0\\[1em]
			0 &\Phi^{m}_1&0\cdots 0&&\cdots&&0 &\phi^{m}_1&0\cdots 0&&\cdots&&0 &\phi^{m}_1&0\cdots 0\\[1em]
			0 &0\cdots0&\Phi^M_1&&\cdots&&0 &0\cdots0&\phi^M_1&&\cdots&&0 &0\cdots0&\phi^M_1\\[1em]
			
			\vdots &\vdots&\vdots&&\ddots&&\vdots&\vdots&\vdots &&\ddots&&\vdots &\vdots&\vdots\\[1em]
			
			\phi_n^{1} &0\cdots0&0&&\cdots&&\Phi_n^{1}&0\cdots0&0&&\cdots&&\phi_n^{1}&0\cdots0&0\\[1em]
			0 &\phi^{m}_n&0\cdots 0&&\cdots&&0 &\Phi^{m}_n&0\cdots 0&&\cdots&&0 &\phi^{m}_n&0\cdots 0\\[1em]
			0 &0\cdots0&\phi^M_n&&\cdots&&0 &0\cdots0&\Phi^M_n&&\cdots&&0 &0\cdots0&\phi^M_n\\[1em]
			
			\vdots &\vdots&\vdots&&\ddots&&\vdots&\vdots&\vdots &&\ddots&&\vdots &\vdots&\vdots\\[1em]
			
			\phi_N^{1} &0\cdots0&0&&\cdots&&\phi_N^{1}&0\cdots0&0&&\cdots&&\Phi_N^{1}&0\cdots0&0\\[1em]
			0 &\phi^{m}_N&0\cdots 0&&\cdots&&0 &\phi^{m}_N&0\cdots 0&&\cdots&&0 &\Phi^{m}_N&0\cdots 0\\[1em]
			0 &0\cdots0&\phi^M_N&&\cdots&&0 &0\cdots0&\phi^M_N&&\cdots&&0 &0\cdots0&\Phi^M_N\\[1em]
			
		\end{bmatrix}.
		\endgroup 
		\label{eq:t2}
	\end{equation}
	\hrule
\end{figure*}
The entries of $\textbf{G}(\mathbf{s},\omega)$ can then be calculated as follows: 

\begin{equation}
	\label{eq:Phi}
	\begin{split}
		\Phi_n^m&=\dfrac{2R_m}{\sum_{n=1}^N B_n}\bigg(\dfrac{-T_m^2}{(s^m_n+T_m)^3}\bigg),
	\end{split}
\end{equation}
and 
\begin{equation}
	\label{eq:phi}
	\phi_n^m=
	\dfrac{2R_m}{\sum_{n=1}^N B_n}\bigg(\dfrac{s^m_nT_m}{(s^m_n+T_m)^3}\bigg).
\end{equation}
From~\eqref{eq:Phi} and~\eqref{eq:phi}, we can calculate
\begin{equation}
	\label{eq:delta}
	\Delta_n^m=\Phi_n^m-\phi_n^m=
	\dfrac{2R_m}{\sum_{n=1}^N B_n}\bigg(\dfrac{-T_m(s^m_n+T_m)}{(s^m_n+T_m)^3}\bigg),
\end{equation}
which is negative.
Then, $\textbf{G}(\mathbf{s},\omega)$ can be expressed as a sum of 2 matrices $\textbf{G}=\textbf{D}+\textbf{E}$, where:
\begin{itemize}
	\item  $\textbf{D}$ is similar to $\textbf{G}$, except that all the diagonal entries of $\textbf{D}$ are $\phi_n^m$ instead of $\Phi_n^m$. Then, $\textbf{D}$ has identical columns (columns $i$ and $M+i$ are identical), and thus it is negative semi-definite.
	\item $\textbf{E}$ is a diagonal matrix with entries equal to $\Delta_n^m$. Thus, $\textbf{E}$ is negative definite 
\end{itemize}
As a result, $\textbf{G}(\mathbf{s},\omega)$ is the sum of a negative semi-definite matrix and a negative definite matrix ($\textbf{E}$). Thus, $\textbf{G}(\mathbf{s},\omega)$ is negative definite. Therefore, $[\textbf{G}(\mathbf{s},\omega)+\textbf{G}^{T}(\mathbf{s},\omega)]$ is negative definite, and the proof is completed.
	\section{Proof of Theorem 5}
\label{proof5}

Assume that follower $n$ is employing strategy $\mathbf{s}_n$ which invests less than the available budget, i.e., $\sum_{m=1}^M~s^m_n~<~B_n$. The utility function in this case is given in~\eqref{eq:totalpayoff}. Without loss of generality, if the follower chooses a strategy $\mathbf{s}'_n$ which invests the remaining budget $\Delta s^j_n$ into a chain $j$, its utility function becomes:
\begin{equation}
	\label{eq:B2}
	\begin{split}
		U'_n=\sum_{m \in \mathcal M_{-j}}\bigg(\dfrac{s_n^m}{s^m_n+T_m}R_m\bigg) + \dfrac{s_n^j+\Delta s^j_n}{s_n^j+\Delta s^j_n+T_j}R_j,
	\end{split}
\end{equation}
where $\mathcal M_{-j}$ is the set of all chains except chain $j$. Then, the difference in the utilities between the two strategies is:
\begin{equation}
	\label{eq:B2}
	\begin{split}
		U'_n-U_n&=\dfrac{s_n^j+\Delta s^j_n}{s_n^j+\Delta s^j_n+T_j}R_j-\dfrac{s_n^j}{s_n^j+T_j}R_j,\\
		&=\dfrac{\Delta s^j_n\sum_{k \in \mathcal{N}_{-n}} s_k^j}{(s_n^j+\Delta s^j_n+T_j)(s_n^j+T_j)},
	\end{split}
\end{equation} 
which is always positive. This means that $\mathbf{s}_n$ always gives a lower payoff than $\mathbf{s}'_n$ regardless of the other followers' strategies, and the proof is completed.
	\section{Proof of Theorem 6}
\label{proof6}
We prove that at the point where every follower's strategy satisfies $s_n^m=B_n\dfrac{R_m}{\sum_{i=1}^{M}R_i}, \forall m \in \mathcal{M}, \forall n \in \mathcal{N}$, every follower's strategy maximizes its utility ($\dfrac{\partial U_n}{\partial s_n^m}=0$). Therefore, no rational follower will deviate from this point, and thus this is the Nash equilibrium of this game. Substitute $s_n^m=B_n\dfrac{R_m}{\sum_{i=1}^{M}R_i}$ into $\dfrac{\partial U_n}{\partial s_n^m}$, we have
\begin{equation}
	\begin{aligned}
		\begin{split}
			\dfrac{\partial U_n}{\partial s_n^m}=&\dfrac{R_{m}T_{m}}{(s^{m}_n+T_{m})^2}-\dfrac{R_MT_M}{(s_n^M+T_M)^2},\\
			=&\dfrac{\sum_{j \in \mathcal{N}_{-n}}B_j\dfrac{R_m^2}{\sum_{i=1}^{M}R_i}}{\sum_{n=1}^{N}(B_n\dfrac{R_m}{\sum_{i=1}^{M}R_i})^2}-\dfrac{\sum_{j \in \mathcal{N}_{-n}}B_j\dfrac{R_M^2}{\sum_{i=1}^{M}R_i}}{\sum_{n=1}^{N}(B_n\dfrac{R_M}{\sum_{i=1}^{M}R_i})^2},\\
			=&\sum_{j \in \mathcal{N}_{-n}}B_j\bigg(\dfrac{\sum_{i=1}^{M}R_i}{\sum_{n=1}^{N}(B_n)^2}-\dfrac{\sum_{i=1}^{M}R_i}{\sum_{n=1}^{N}(B_n)^2}\bigg),\\
			=&0, \forall m \in \mathcal{M},\text{and }  \forall n\in \mathcal{N}.
		\end{split}
	\end{aligned}
	\label{eq:t43}
\end{equation}
The proof is now completed.
	\section{Proof of Theorem 7}
\label{proof7}
To find the equilibrium of this upper sub-game, we first find the best response $R^*_m$ for each leader, i.e., the strategies that maximizes $U_m$ when the strategies of the other leaders are fixed. To this end, we first take the derivative of $U_m$:
\begin{equation}
	\label{eq:leader}
	\begin{split}
\dfrac{\textrm{d} U_m}{\textrm{d} R_m}=\sum_{n=1}^N\dfrac{B_n\sum_{i \in \mathcal{M}_{-m}}R_i\bigg(1+\lnb{\dfrac{B_nR_m}{\sum_{i=1}^{M}R_i}}\bigg)}{(R_m+\sum_{i \in \mathcal{M}_{-m}}R_i)\big)^2}-1.
	\end{split}
\end{equation} 

To find $R^*_m$, we solve $\dfrac{\textrm{d} U_m}{\textrm{d} R_m}=0$, i.e., 
\begin{equation}
	\label{eq:leader2}
	\begin{split}
			\sum_{n=1}^N\dfrac{B_n\sum_{i \in \mathcal{M}_{-m}}R_i\bigg(1+\lnb{\dfrac{B_nR_m}{\sum_{i=1}^{M}R_i}}\bigg)}{(R_m+\sum_{i \in \mathcal{M}_{-m}}R_i)\big)^2}-1=0.
	\end{split}
\end{equation} 
Since the leaders' utility functions are the same, we have $\sum_{i \in \mathcal{M}_{-m}}R_i=(M-1)R_m$. Then,~\eqref{eq:leader2} becomes
\begin{equation}
	\label{eq:leader}
	\begin{split}
		\dfrac{\partial U_m}{\partial R_m}=\sum_{n=1}^N\dfrac{B_n\sum_{i \in \mathcal{M}_{-m}}R_i\bigg(1+\lnb{\dfrac{B_nR_m}{\sum_{i=1}^{M}R_i}}\bigg)}{(R_m+\sum_{i \in \mathcal{M}_{-m}}R_i)^2}-1&=0,\\
		\sum_{n=1}^N\dfrac{B_n\sum_{i \in \mathcal{M}_{-m}}R_i\bigg(1+\lnb{\dfrac{B_nR_m}{\sum_{i=1}^{M}R_i}}\bigg)}{(R_m+\sum_{i \in \mathcal{M}_{-m}}R_i)^2}&=1,\\
		\sum_{n=1}^N\dfrac{B_n(M-1)R_m\bigg(1+\lnb{\dfrac{B_nR_m}{MR_m}}\bigg)}{(MR_m)^2}&=1,\\
		\sum_{n=1}^N\dfrac{B_n(M-1)\bigg(1+\lnb{\dfrac{B_n}{M}}\bigg)}{M^2R_m}&=1,\\
		\dfrac{M-1}{M^2}\sum_{n=1}^NB_n\bigg(1+\lnb{\dfrac{B_n}{M}}\bigg)&=R_m.\\
	\end{split}
\end{equation} 
Thus,
\begin{equation}
	\label{eq:optimalleader}
	R^*_m=\dfrac{M-1}{M^2}\sum_{n=1}^NB_n\bigg(1+\lnb{\dfrac{B_n}{M}}\bigg),
\end{equation} 
is the optimal strategy of leader $m$. Since $R^*_m$ is uniquely defined by constants, i.e., $M$ and $B_n$, the equilibrium of this upper sub-game exists and is unique. As a result, the considered Stackelberg game admits a unique Stackelberg equilibrium.

\section{Algorithm 1}
\label{proof8}
\begin{algorithm}[H]
	\caption{Simulation Steps}\label{euclid}
	\begin{algorithmic}[1]
		\State $k \gets 0$
		\Repeat
		\If{reward scheme = dynamic} \LineComment{Chains set block rewards at each epoch}
		\For {$m:=1$ to $M$}
		\State $R^*_m\gets \dfrac{M-1}{M^2}\sum_{n=1}^NB_n\bigg(1+\lnb{\dfrac{B_n}{M}}\bigg)$
		\EndFor
		\EndIf
		\For {$n:=1$ to $N$}\LineComment{Followers make decisions}
		\For {$m:=1$ to $M$}	 		
		\State Find $s^{*m}_n$ using fmincon	
		\EndFor 
		\EndFor 
		\If{Adversary = Static}\LineComment{Static Adversary}
		\State Adversary attacks with fixed $B_{A}$
		\Else\LineComment{Adaptive Adversary}
		\State Adversary corrupts $N_A$ stakeholders
		\State Adversary attacks with $B_{A}=\sum_{i \in \mathcal{N}_{A}} B_i$ 
		\EndIf
		\For {$i:=1$ to $N_\Delta$}\LineComment{Randomly adjust followers' budgets}	 		
		\State Adjust a random follower budget by $\pm \Delta_sB_n$	
		\EndFor   		
		\State $k \gets k+1$
		\Until {$k>n_e$}
	\end{algorithmic}
\end{algorithm}
\end{document}